\documentclass[twocolumn,showpacs,preprintnumbers,superscriptaddress,amsmath,amssymb]{revtex4}
\usepackage{graphicx}% Include figure files
\usepackage{dcolumn}% Align table columns on decimal point
\usepackage{bm}% bold math
\usepackage{amssymb}
\usepackage{amssymb}
\usepackage{verbatim}
\usepackage{color}
\def\>{\rangle}

\makeatletter

\newcommand{\Rmnum}[1]{\expandafter\@slowromancap\romannumeral #1@}
\makeatother

\begin{document}

\title{Synchronization in $\mathcal{PT}$-symmetric optomechanical resonators}% Force line breaks with\\
\author{Chang-long Zhu}
\affiliation{Department of Automation, Tsinghua University,
Beijing 100084, P. R. China}
\author{Yu-long Liu}
\affiliation{Institute of Microelectronics, Tsinghua University,
Beijing 100084, P. R. China}
\author{Lan Yang}
\affiliation{Department of Electrical
and Systems Engineering, Washington University, St.~Louis, MO
63130, USA}
\author{Yu-xi Liu}\email{yuxiliu@mail.tsinghua.edu.cn}
\affiliation{Institute of Microelectronics, Tsinghua University,
Beijing 100084, P. R. China} \affiliation{Center for Quantum
Information Science and Technology, BNRist, Beijing 100084, P. R.
China}
\author{Jing Zhang}\email{jing-zhang@mail.tsinghua.edu.cn}
\affiliation{Department of Automation, Tsinghua University,
Beijing 100084, P. R. China} \affiliation{Center for Quantum
Information Science and Technology, BNRist, Beijing 100084, P. R.
China}

\date{\today}

\begin{abstract}
Synchronization has great impacts in various fields such as
self-clocking, communication, neural networks, etc. Here we
present a mechanism of synchronization for two mechanical modes in
two coupled optomechanical resonators via optical coupling
of the cavity fields by introducing the so-called
$\mathcal{PT}$-symmetric structure. It is shown that the degree of
synchronization between the two far-off-resonant mechanical modes can be increased
by decreasing the coupling strength between the two optomechanical resonators.
Furthermore, when we consider the stochastic noises in the
optomechanical resonators, we find that more noises can enhance
the degree of synchronization of the system under particular
$\mathcal{PT}$-symmetric condition. Our results reveal versatile
effects of optical $\mathcal{PT}$-symmetry on controlling the
synchronization dynamics of indirect coupled mechanical resonators.
parameter regime.
\end{abstract}


\maketitle

\section{Introduction}\label{s1}
Synchronization is a phenomenon in which two or more systems coordinate
and act at the same time with similar behavior. Synchronization determined
phenomenon such as the chorusing of crickets, a flash of fireflies, pendulum
clocks, and even the life cycle of creatures~\cite{Pikovsky,RBrown,Glass}
have been extensively observed in our daily life. In particular, synchronization,
the rhythms of two or more different objects adjusted in unison, is a qualitative
transition and thus motives wide applications in various fields, such as data
communication, timekeeping, navigation, cryptography, and neuroscience~\cite{Winfree,Goldbeter,Taylor,Strogatz,Manrubia,Bregni}.

Benefiting from current advanced nano-fabrication techniques, especially
those for high-quality-factor on-chip optomechanical
resonators~\cite{Aspelmeyer}, it is possible to demonstrate the synchronization
of resonators in on-chip nano-scale platforms~\cite{Holmes,Li,Zhangmian,Bagheri,Shah,YangNan}.
For example, a pair of closely placed optomechanical resonators with different
mechanical frequencies were synchronized by indirect coupling through the coupled
optical fields~\cite{Zhangmian}. More recently, two nanomechanical oscillators
separated for about $80$ $\mu$m were synchronized through the same optical field
in an optical racetrack~\cite{Bagheri}.

In this paper, we show that mechanical oscillations can be
synchronized by optomechanical couplings to two coupled optical
modes, in which one is active and the other one is passive.
With balanced gain and loss, such kinds of systems are called
parity-time~($\mathcal{PT}$)-symmetric optomechanical systems,
which have attracted great attentions in recent
years~\cite{HXu,Jinghui,Schonleber,Liuzhongpeng,Jinghui2,Zhangjing1,Lvxinyou}.
Various appealing phenomena and important applications have
been proposed in particular systems with $\mathcal{PT}$-symmetric
structure~\cite{HXu,Jinghui,Schonleber,Liuzhongpeng,Jinghui2,Zhangjing1,Lvxinyou,Bender1,Bender2,Agarwal,Mostafazadeh,Pengbo,Feng1,Hodaei,Guo,Ruter,Ramezani,Lin,Feng2,Regensburger,Changlong,Pengbo2,Schindler,West,JWiersig,WChen,HHodaei,JDoppler}.

Although the optomechanical interaction has influence
on our $\mathcal{PT}$-symmetric system, this influence is negligibly small
under the parameter regime we consider~\cite{Jinghui,Schonleber,Liuzhongpeng,Jinghui2,Lvxinyou}.
By introducing the $\mathcal{PT}$-symmetric structure, we observe
an interesting phenomenon that the two mechanical modes of the
coupled optomechanical resonators tend to oscillate in unison by
decreasing the optical coupling strength between them. This
observation somewhat conflicts with the normal phenomenon that:
the stronger coupling strength between two systems is, the easier
the synchronization can be realized. Another counterintuitive
phenomenon presented as the enhancement of synchronization between
the two mechanical modes when considering the noises acting on
the optomechanical resonators.

\section{Coupled-optomechanical resonators with optical $\mathcal{PT}$-symmetry}

\begin{figure}[h]
\centerline{
\includegraphics[width=8.4 cm, clip]{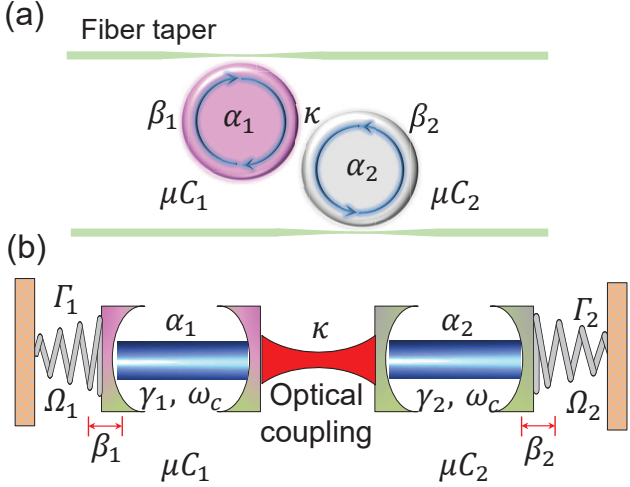}}
\caption{(color online) Schematic diagram of the optically-coupled
$\mathcal{PT}$ optomechanical system. (a) $\mu C_1$ denotes an
active WGM resonator with gain medium and $\mu C_2$ is a passive
one. (b) Equivalent diagram of the $\mathcal{PT}$
optomechanical system, where the WGM resonators are replaced by
Fabry-Perot cavities with a moveable end mirror and a fixed one.
The two cavities are directly coupled through the inter-cavity evanescent optical
fields, and the optical coupling strength $\kappa$ depends on
the distance between the two Fabry-Perot cavities~\cite{Zhangmian}. }\label{Fig1}
\end{figure}

The system we consider consists of two coupled whispering-gallery-mode (WGM)
resonators, and is depicted in Fig.~\ref{Fig1}(a). The left WGM
resonator~($\mu C_1$) is an active one which can be realized,
e.g., by $\rm{Er}^{3+}$-doped silica disk, and the right one~($\mu
C_2$) is a passive resonator. Each resonator supports an optical
mode $\alpha_i$ and a mechanical mode $\beta_i$ ($i=1,2$), and the
inter-cavity optical coupling strength $\kappa$ between $\alpha_1$
and $\alpha_2$ is related to the distance between the two resonators.
As is well known, although the two mechanical modes
$\beta_1$ and $\beta_2$, located in two different resonators, are
not directly coupled, they can be indirectly
coupled through the inter-cavity optical coupling and the
intra-cavity optomechanical coupling. We elaborate this indirect
mechanical coupling in Fig.~\ref{Fig1}(b). Each WGM resonator is
equivalent to a Fabry-Perot cavity, with one fixed mirror and one
movable one. The optical modes $\alpha_1$ and $\alpha_2$ represent
the optical fields in the Fabry-Perot cavities and the mechanical
modes $\beta_1$ and $\beta_2$ indicate the motions of the movable
mirrors. In each equivalent Fabry-Perot cavity, the movable mirror
suffers a radiation-pressure force induced by the optical mode
$\alpha_i$ (i=1,2). Such a force is proportional to the
circulating optical intensity $|\alpha_i|^2$ in the cavity, which
leads to the mechanical motion $\beta_i$. In the meantime, the
movable mirror induces a frequency-shift of the optical mode in
the cavity, which influences the dynamics of $\alpha_i$. In
Fig.~\ref{Fig1}(b), $\alpha_1$ ($\alpha_2$) and $\beta_1$
($\beta_2$) interact with each other through this kind of
radiation-pressure coupling, and $\alpha_1$ and $\alpha_2$ are directly
coupled through the inter-cavity evanescent optical fields. Therefore, the
mechanical modes $\beta_1$ and $\beta_2$ are coupled indirectly by
the evanescent optical coupling between $\alpha_1$ and $\alpha_2$.

The $\mathcal{PT}$-optomechanical system we consider can be
represented by the following equations:
\begin{eqnarray}\label{Dynamical Equations of the optomechanical system}
\dot{\alpha}_1&=&-\Gamma_{\rm op1}\alpha_1-i \kappa \alpha_2 -i
g_{om}\alpha_1(\beta_1+\beta^*_1)
+\sqrt{2\gamma_{1ex}}\epsilon_1,\nonumber\\
\dot{\alpha}_2&=&-\Gamma_{\rm op2}\alpha_2-i \kappa \alpha_1 -i
g_{om}\alpha_2(\beta_2+\beta^*_2)
+\sqrt{2\gamma_{2ex}}\epsilon_2,\nonumber\\
\dot{\beta}_1&=&-(\Gamma_{m1}+i\Omega_1)\beta_1-i g_{om} |\alpha_1|^2,\nonumber\\
\dot{\beta}_2&=&-(\Gamma_{m2}+i\Omega_2)\beta_2-i g_{om}
|\alpha_2|^2,
\end{eqnarray}
where $\Gamma_{\rm op1}=-\gamma_1+i\Delta_1$ and $\Gamma_{\rm
op2}=\gamma_2+i\Delta_2$. $\gamma_i$, $\gamma_{iex}$, $\Delta_i=\omega_{ci}-\omega_{L}$,
and $\epsilon_i$ ($i=1,2$) denote the gain (loss) rate of the
resonator $\mu C_i$, the external damping rate induced by the
coupling between the resonator and the input/output fiber-taper,
the detuning frequency between the resonance frequency ($\omega_{ci}$)
of the cavity mode and the frequency ($\omega_{L}$) of the driving
field, and the amplitude of the driving field, respectively.
Without loss of generality, here we assume that $\Omega_2 \ge \Omega_1$.
$\Omega_i$ and $\Gamma_{mi}$ represent the frequency and damping
rate of the mechanical mode $\beta_i$.
To simplify our discussion, we assume that the gain cavity
$\mu C_1$ and the lossy cavity $\mu C_2$ have the same
vacuum optomechanical coupling strength $g_{om}$
which quantifies the interaction between a single photon and a single phonon.
We also assume that the gain rate of $\mu C_1$ is equal to the
damping rate of $\mu C_2$, i.e., $\gamma_2=\gamma_1\equiv\gamma$,
which means that the gain and loss in the system are well
balanced. Additionally, we consider the case of critical coupling
such that $\gamma_{1ex}=\gamma_{2ex}=\gamma/2$.

In general, the vacuum optomechanical coupling strength $g_{om}$ of
typical optical cavities is very small~\cite{Aspelmeyer}, and thus that
the influence of optomechanical interaction on optical structure in our
system can be ignored. Under the condition of symmetric optical driving
detunings ($\Delta_{-}=\Delta_2-\Delta_1=0$), there exists a phase
transition point, called exceptional point (EP)~\cite{Jinghui,Schonleber,Liuzhongpeng,Jinghui2,Lvxinyou},
corresponding to a critical inter-cavity coupling strength
$\kappa_{\rm EP}=\gamma$. When $\kappa>\kappa_{\rm EP}$ which
is in so-called $\mathcal{PT}$-symmetric regime, there
exist two non-degenerate optical supermodes with the same damping
rate. When $\kappa\le\kappa_{\rm EP}$ which is in the so-called broken
$\mathcal{PT}$-symmetric regime, the two optical supermodes
are degenerate but with different damping rates.
When the system is far away from the EP, the
interaction between the optical supermodes and mechanical modes,
i.e. the effective radiation-pressure coupling in the supermode
picture, is weak. This kind of interaction will be
greatly enhanced as $\kappa$ approaches to $\kappa_{\rm EP}$. This
results from the topological-singularity-induced amplification
of the optomechanical nonlinearity in the vicinity of the exceptional
point~\cite{Jinghui,Schonleber,Liuzhongpeng,Jinghui2,Zhangjing1}.

However, slightly different from Refs~\cite{Jinghui,Schonleber,Liuzhongpeng,Jinghui2,Lvxinyou},
in this work we consider asymmetric optical driving detunings, i.e., $\Delta_{-}=\Delta_2-\Delta_1\ne 0$,
in order to synchronize the two mechanical modes which will be discussed in the following
section. The difference between the two optical driving detunings $\Delta_{-}$
is small enough that the properties of $\mathcal{PT}$-symmetric structure
in our system is still held, i.e., the optomechanical
interaction can still be greatly amplified near the exceptional point.
Here, we consider the condition (see Appendix \ref{Weaker condition of PT-symmetry})
\begin{equation}\label{Weaker condition of PT-symmetric optomechanical system}
g_{om} \ll \Delta_{-}\ll \sqrt[3]{\frac{2}{3}\gamma\left(g_{om}^{2}\frac{\Omega_2+\Omega_1}{\Omega_1\Omega_2}\gamma\epsilon^2 \right)^{2}} \ll \gamma, \kappa,
\end{equation}
then the non-degeneracy between the optical supermodes at exceptional point
can be approximated given by
\begin{equation}
\frac{\Delta_{\rm{split}}}{\gamma}\approx \sqrt{\frac{\Delta_{-}^3}{\frac{2}{3}\gamma
\left( g_{om}^2\frac{\Omega_2+\Omega_1}{\Omega_1\Omega_2} \gamma\epsilon^2\right)^{2/3}}},\nonumber
\end{equation}
where $\Delta_{\rm{split}}=\rm{Im}[\omega_{o+}-\omega_{0-}]=\rm{Re}[\omega_{o+}-\omega_{0-}]$,
and $\omega_{o\pm}$ are the eigenvalues of optical supermodes.
It is clear that this non-degeneracy $\Delta_{\rm{split}}$ is very small
that the $\mathcal{PT}$-symmetric structure in our system is still held.

\begin{figure}[h]
\centerline{
\includegraphics[width=8.6 cm,clip]{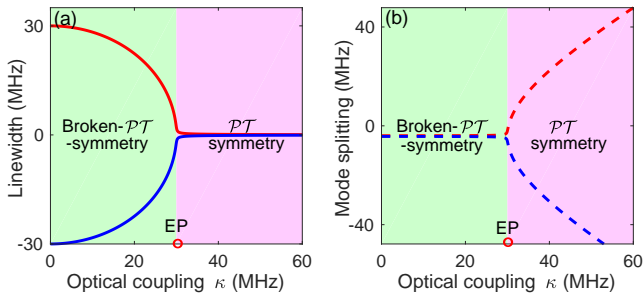}}
\caption{(Color online) (a) Linewidth of the supermodes, i.e., the
real parts of the eigenfrequencies, (b) mode splitting of the
supermodes, i.e., the imaginary parts of the eigenfrequencies.
The green region is the broken-$\mathcal{PT}$-symmetry
regime, the pink region corresponds to the $\mathcal{PT}$-symmetry
regime.}\label{Fig2}
\end{figure}

Given the system parameters $\gamma=30$~MHz, $\Delta_1=4.2$~MHz,
$\Delta_2=5$~MHz, $\Omega_1=5$~MHz, $\Omega_2=15$~MHz, $\Gamma_{m1}=8$~kHz,
$\Gamma_{m2}=8$~kHz, $g_{om}=3$~kHz, and $\epsilon=70$~MHz$^{1/2}$,
the simulation results of the mode splitting and linewidth of the optical
supermodes are shown in Figs.~\ref{Fig2} (a) and (b).
It is obvious that the non-degeneracy at EP in Fig.~\ref{Fig2} is negligibly
small, and the broken-$\mathcal{PT}$-symmetric and $\mathcal{PT}$-symmetric
regimes can be clearly observed. It should be noted that
although one eigenfrequency of the optical supermodes
has the positive real component in the broken-$\mathcal{PT}$-symmetric
regime (Fig.~\ref{Fig2}(a)), the saturation nonlinearity induced by the optomechanical
coupling will suppress the divergence induced by this positive rate~\cite{XinZhou, Hassan}.

\section{Frequency synchronization via $\mathcal{PT}$-symmetry}\label{s3}
When the degrees of freedom of the optical modes are adiabatically eliminated
under the condition that the optical decay rates are much larger than
the mechanical decay rates, the enhanced optomechanical coupling,
induced by the topological-singularity-induced amplification
of the optomechanical nonlinearity, will lead to significant
effective frequency shifts $\delta\Omega_1$ and
$\delta\Omega_2$ for the mechanical modes $\beta_1$ and $\beta_2$
in the vicinity of EP. In fact, under the condition depicted in
Eq.~(\ref{Weaker condition of PT-symmetric optomechanical system})
and $\epsilon_1=\epsilon_2\equiv\epsilon$, $\delta\Omega_1$ and
$\delta\Omega_2$ near EP can be written as (detailed derivation see Appendix~\ref{Appendix: effective frequency shifts and coupling})
\begin{equation}\label{Mechanical frquency shift in the vicinity of exceptional point}
\delta\Omega_{1}=-\delta\Omega_2\approx\frac{g_{om}^2\Delta_{-}(\gamma^2+\kappa^2)^2\gamma\epsilon^2}
{\left[(\kappa^2-\gamma^2)^2+\gamma^2\Delta_{-}^{2} \right]^2}.
\end{equation}
Here, in order to synchronize the two mechanical oscillators, we require that
$\Delta_{1}$ and $\Delta_2$ have small difference, which
makes sure that $\delta\Omega_{1}$ and $\delta\Omega_{2}$ are opposite in sign,
and the influence on the structure of $\mathcal{PT}$-symmetry is very small simultaneously.
\begin{figure}[h]
\centerline{
\includegraphics[width=8.6 cm,clip]{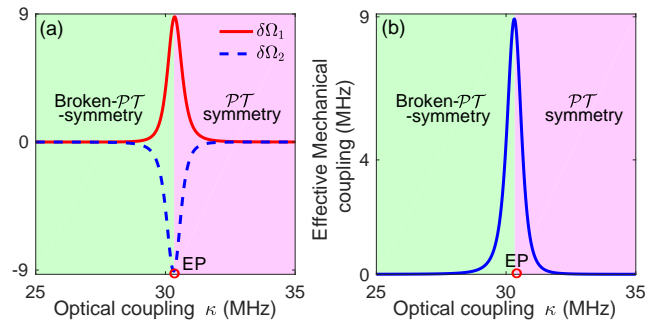}}
\caption{(Color online) (a) Optomechanics-induced mechanical frequency
shifts $\delta\Omega_{1,2}$ of the two optomechanical resonators versus
the optical coupling strength $\kappa$ both in broken-$\mathcal{PT}$-symmetric
regime and $\mathcal{PT}$-symmetric regime.
(b) Effective coupling strength $\kappa_{\rm{mech}}$ between two mechanical
modes versus the optical coupling strength $\kappa$.
}\label{Fig3}
\end{figure}

We show in Fig.~\ref{Fig3}(a) the optomechanics-induced mechanical frequency shift
$\delta\Omega_1$ (red-solid curve) and $\delta\Omega_2$ (blue-dashed curve)
of the two resonators versus the optical coupling strength $\kappa$,
both in broken-$\mathcal{PT}$-symmetric and $\mathcal{PT}$-symmetric regimes.
When the system is far away from the exceptional point, the optomechanics-induced
mechanical frequency shift $\delta\Omega_i$ is negligibly small. However,
$\delta\Omega_i$ will be greatly enhanced such that $\delta\Omega_i$ is
almost comparable with or even larger than $\Omega_i$, when $\kappa$
approaches to $\kappa_{\rm{EP}}$. As these two enhanced frequency shifts for
the mechanical modes are opposite in sign, they will lead to significant
modifications of mechanical frequencies $\Omega_1$ and $\Omega_2$, and
make the two mechanical frequencies $\Omega_1$ and $\Omega_2$ to approach each other.
Thus the two oscillators tend to be resonant with each other,
and occurs synchronization.

Moreover, the enhanced optomechanical coupling can also induce an enhancement
of the effective mechanical interaction between mechanical modes $\beta_1$
and $\beta_2$ in the vicinity of the EP. In fact, by adiabatically eliminating
the degrees of freedom of the optical modes, we obtain the effective coupling
strength $\kappa_{\rm{mech}}$ between the two mechanical modes $\beta_1$ and
$\beta_2$ as
\begin{equation}\label{Effective mechanical coupling in the vicinity of exceptional point}
\kappa_{\rm mech}\approx\frac{4g_{om}^2\Delta_{-}\kappa^2\gamma^3\epsilon^2}
{\left[ (\kappa^2-\gamma^2)^2+\gamma^2\Delta_{-}^2\right]^2}.
\end{equation}
In Fig.~\ref{Fig3}(b) the effective mechanical coupling strength
$\kappa_{\rm{mech}}$ versus the optical coupling strength $\kappa$ is plotted, both
in broken-$\mathcal{PT}$-symmetric and $\mathcal{PT}$-symmetric regimes.
It can be clearly seen that the effective mechanical coupling strength
$\kappa_{\rm{mech}}$ is negligibly small when the system is far away from the
exceptional point, but can be significantly enhanced when $\kappa$ approaches
to $\kappa_{\rm{EP}}$. This enhanced effective mechanical interaction in the vicinity of the EP can also
contribute to synchronization between the two mechanical modes $\beta_1$
and $\beta_2$, since the enhanced $\kappa_{\rm{mech}}$ can greatly change the mechanical frequencies
$\Omega_1$ and $\Omega_2$ and make the two mechanical frequencies to get
close to each other (detailed discussion can be found in
Appendix~\ref{The influence of the effective mechanical coupling on synchronization}).

\begin{figure}[h]
\centerline{
\includegraphics[width=9 cm, clip]{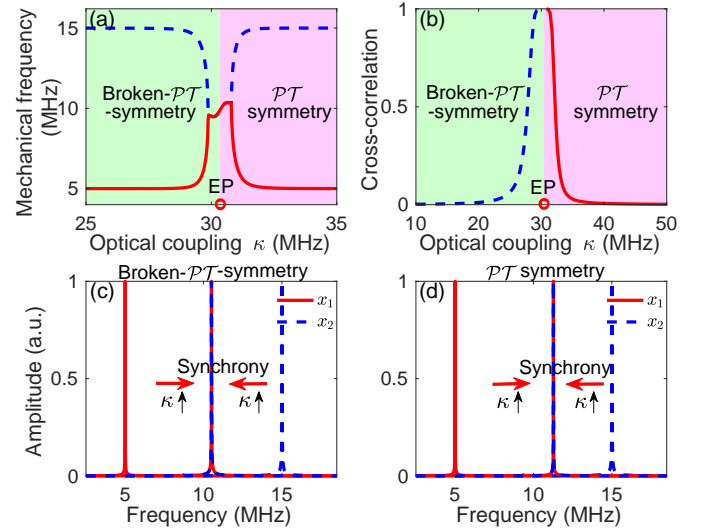}}
\caption{(color online)
(a) Effective mechanical frequencies
$\Omega_{\rm 1,eff}$ and $\Omega_{\rm 2,eff}$ versus the optical
coupling strength $\kappa$, where the red solid (blue dashed)
curve represents the frequency of $\beta_1$ ($\beta_2$), the light
green (pink) area is the broken- $\mathcal{PT}$-symmetric
($\mathcal{PT}$-symmetric) regime.
(b) Numerical results of cross-correlation
$M_{cc}$ with different values of $\kappa$ in broken-$\mathcal{PT}$-symmetric
and $\mathcal{PT}$-symmetric regimes.
(c) Spectrograms of mechanical
modes $x_1$ and $x_2$ with increasing optical coupling strength
$\kappa$ in the broken- $\mathcal{PT}$-symmetric regime.
Here, $\kappa\uparrow$ and $\kappa\downarrow$ denote the increase
and decrease of $\kappa$.
(d) Spectrograms of mechanical modes $x_1$ and $x_2$ with decreasing
optical coupling strength $\kappa$ in the $\mathcal{PT}$-symmetric
regime, in which weaker coupling strength $\kappa$ makes the two
resonators more easily to be synchronized.}\label{Fig4}
\end{figure}

Actually, the effective mechanical frequencies of the two mechanical
oscillators in the vicinity of the EP
can be expressed as $\Omega_{\rm{1,eff}}=\Omega_1+\delta\Omega_1+\delta\Omega_{\rm{coup}}$
and $\Omega_{\rm{2,eff}}=\Omega_2+\delta\Omega_2-\delta\Omega_{\rm{coup}}$,
respectively, where $\delta\Omega_{\rm{coup}}$ is induced by the effective
mechanical coupling strength $\kappa_{\rm{mech}}$
(see Appendix~\ref{The influence of the effective mechanical coupling on synchronization}).
This means that the enhanced optomechanics-induced mechanical frequency
shifts $\delta\Omega_1 / \delta\Omega_2$ and effective mechanical coupling strength $\kappa_{\rm{mech}}$
can result in significant modifications of mechanical frequencies $\Omega_1 / \Omega_2$ together,
and thus jointly contribute to the synchronization between the two mechanical oscillators,
i.e., $\Omega_{\rm{1,eff}}=\Omega_{\rm{2,eff}}$. We show in Fig.~\ref{Fig4}(a) that the
effective mechanical frequencies $\Omega_{\rm{1,eff}}$ (red-solid curve)
and $\Omega_{\rm{2,eff}}$ (blue-dashed curve) of the two resonators versus
the optical coupling strength $\kappa$, both in broken-$\mathcal{PT}$-symmetric
and $\mathcal{PT}$-symmetric regimes. It is clear that the two mechanical
oscillators tend to be resonant with each other, i.e., $\Omega_{\rm{1,eff}}=\Omega_{\rm{2,eff}}$,
and thus synchronize, when $\kappa$ approaches to $\kappa_{\rm{EP}}$.
As is well known, the frequency-mismatch between two synchronized oscillators should
be very small in traditional lossy systems~\cite{Li,Zhangmian}, i.e., $|\Omega_1-\Omega_2|\ll
\Omega_{1},\Omega_2$. However, as shown in Fig.~\ref{Fig4}, our $\mathcal{PT}$-symmetric
system can perfectly synchronize two far-off-resonant mechanical oscillators.
Actually, as shown in Fig.~\ref{Fig4}(a), the effective mechanical frequencies of the
two optomechanical resonators $\Omega_{\rm{1,eff}}$ and $\Omega_{\rm{2,eff}}$
coincide with each other when $\kappa$ approaches $\kappa_{\rm{EP}}$.

In addition, we find an counterintuitive phenomenon that {\it
weaker} coupling between two optomechanical resonators may be {\it
helpful} for synchronization for our $\mathcal PT$ optomechanical
system. In fact, as shown in Fig.~\ref{Fig4}(a), in the $\mathcal
PT$-symmetric regime (the pink region), when the
coupling strength $\kappa$ between two resonators is {\it decreased}, the effective
mechanical frequencies of the two resonators tend to coincide with
each other, which means that $\beta_1$ and $\beta_2$ are inclined
to oscillate in unison with the weaker coupling strength $\kappa$ in
the $\mathcal PT$-symmetric regime. The broken-$\mathcal
PT$-symmetric regime is the normal regime where stronger coupling
between the two optomechanical resonators makes the two mechanical
modes $\beta_1$ and $\beta_2$ be inclined to be synchronized. We
can more easily see this phenomenon by plotting the spectra of the
normalized mechanical displacements of the two optomechanical
resonators $x_1=(\beta_1+\beta_1^*)/2$ (the red solid curve) and
$x_2=(\beta_2+\beta_2^*)/2$ (the blue dashed curve) in
Figs.~\ref{Fig4} (c) and (d), where $\kappa$ is increased from $2$ MHz
to $29.86$ MHz in Fig.~\ref{Fig4} (c), and is decreased from $50$ MHz to
$30.81$ MHz in Fig.~\ref{Fig4} (d).

To give more insights into the phenomena shown by us, we plot in
Fig.~\ref{Fig4}(b) the cross-correlation function $M_{cc}$ between
the two mechanical displacements $x_1$ and $x_2$ with different
inter-cavity optical coupling strength $\kappa$, where $M_{cc}$ is
defined as~\cite{RNBracewell,Rabiner,Anstey,White,Heel,Lewis}
\begin{eqnarray}
M_{cc}&=&\mathop{\max}\limits_{0<t<+\infty}\frac{1}{\sqrt{\phi_1\phi_2}}{\int_{0}^{+\infty}
{x_1(\tau-t)x_2(\tau)d\tau}},\nonumber\\
\phi_i&=&\int_{0}^{+\infty}{x_i^2(\tau)d\tau}.
\end{eqnarray}
This normalized cross-correlation function varies between
0 and 1. The maximum value of $M_{cc}=1$ indicates that the two time
series of the mechanical displacements $x_1$ and $x_2$ have the exact same shape,
even though their amplitudes may be different, which implies that the
two self-sustained oscillators have the same frequency, that is, the
onset of synchronization.
As shown in Fig.~\ref{Fig4}(b), in the $\mathcal{PT}$-symmetric
regime, smaller $\kappa$ induces higher value of $M_{cc}$ (the red
solid curve), and $M_{cc}$ reaches the maximum value (the unit) as
$\kappa$ decreases and approaches EP, which means that the two
mechanical displacements $x_1$ and $x_2$ tend to be synchronized
with the decrease of the inter-cavity coupling strength. However, in the
broken-$\mathcal{PT}$ symmetric regime (the blue dashed curve),
the cross-correlation function increases and tends to unit with
the increase of $\kappa$, which means that stronger inter-cavity
coupling strength will be helpful for synchronization as we
expect.

\section{Noise-enhanced synchronization in $\mathcal{PT}$-symmetric optomechanical system}\label{s4}

\subsection{Stochastic noises in the optical modes}
We now study the effects of the stochastic noises on our
$\mathcal{PT}$-symmetric system. Two
independently-identically-distributed Gaussian white noises
$\xi_{1,2}$ are introduced for the two optical modes
$\alpha_{1,2}$, such that $\left<\xi_i(t)\,\xi_j(t+\tau)\right>=2D\delta_{ij}\delta(\tau)$,
where $D$ is the intensity of the noises. Here, we
have included the shifts of damping rates induced by stochastic noises
into the gain ($\gamma_1$) and loss ($\gamma_2$) rates in our optomechanical system.
Thus the dynamical equations of our $\mathcal{PT}$-symmetric system can be
reexpressed as
\begin{eqnarray}\label{Dynamical Equations with noises}
\dot{\alpha}_1&=&i\left(\Delta_1+g_{om}x_1\right)\alpha_1+\gamma_1\alpha_1-i\kappa \alpha_2 +\sqrt{2\gamma_{1ex}}\epsilon_1 \nonumber\\
&&+\xi_1(t),\nonumber\\
\dot{\alpha}_2&=&i\left(\Delta_2+g_{om}x_2\right)\alpha_2-\gamma_2\alpha_2-i\kappa \alpha_1 +\sqrt{2\gamma_{2ex}}\epsilon_2  \nonumber\\
&&+\xi_2(t),\nonumber\\
\ddot{x}_1&=&-2\Gamma_{m1}\dot{x}_1-\Omega_1^2x_1-g_{om}\left|\alpha_1\right|^2,\nonumber\\
\ddot{x}_2&=&-2\Gamma_{m2}\dot{x}_2-\Omega_2^2x_2-g_{om}\left|\alpha_2\right|^2.
\end{eqnarray}

\begin{figure}[ptb]
\centerline{ \includegraphics[width=8.5 cm, clip]{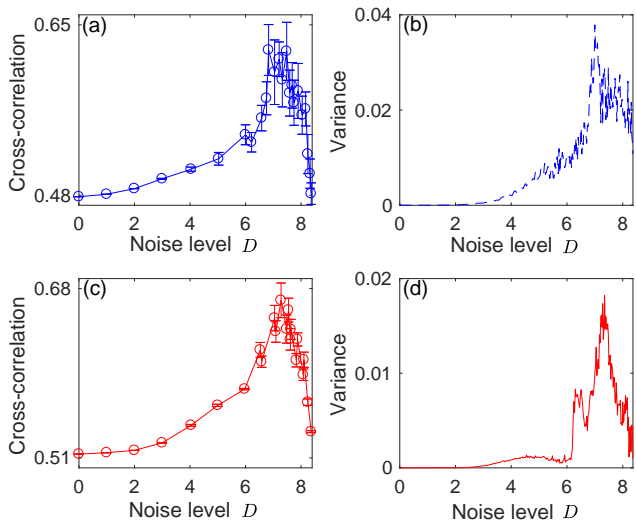}}
\caption{(color online) (a)Effects of the stochastic noises on $M_{cc}$ with
respect to different stochastic noise intensity $D$ in broken
$\mathcal{PT}$-symmetric regime with $\kappa=27.76$ MHz.
(b) Variances of $M_{cc}$ versus noise level $D$ in (a).
(c) Effects of the stochastic noises on $M_{cc}$ versus different $D$ in
$\mathcal{PT}$-symmetric regime with $\kappa=32.19$ MHz. The variance of
$M_{cc}$ is presented in (d).
}\label{Fig5}
\end{figure}

We present the numerical results of the cross-correlation function $M_{cc}$ between the two
mechanical oscillators in Figs.~\ref{Fig5}(a) and (c) by changing
the noise strength $D$ and fixing other parameters both in
broken-$\mathcal{PT}$-symmetric and $\mathcal{PT}$ -symmetric
regimes. It can be seen that $M_{cc}$ is enhanced with increasing
noise intensity $D$ both in broken-$\mathcal{PT}$-symmetric and
$\mathcal{PT}$-symmetric regime, reaches the maximal values at
particular noise level, and then decreases at higher noise
intensity. It means that synchronization process may benefit from
noises~\cite{Neiman1,Han,Neiman2,Nakao,Nagai,Lai,Zhou,Daihai} in
our optomechanical $\mathcal{PT}$-symmetric system. To interpret
what we observe, we can see that the noise will randomly shift the
frequencies of the mechanical modes, especially when we approach
the EP where the effects of noise are
enhanced~\cite{HSchomerus2,SYLee,GYoo,JZhang}. Since the
frequencies of the two mechanical modes are far-separated, these
random frequency shifts may decrease the difference between the
frequencies of the two mechanical modes in a certain probability
with increasing noise strength $D$, and thus increase the
cross-correlation function $M_{cc}$. When we increase the noise
strength $D$ further, the noise will be strong enough to destroy
the periodic oscillation of single mechanical oscillator and the
$\mathcal{PT}$-symmetric structure of the optomechanical system, and thus
decrease the degree of synchronization between the two mechanical
oscillators. This interpretation can also be confirmed by checking
the variance of $M_{cc}$ versus the noise strength $D$ (
Fig.~\ref{Fig5}(b) and (d)). The variance of $M_{cc}$ first
increases with increasing noise strength $D$ (note that $M_{cc}$
increases at the same time), which means that more noises enter
the system although $M_{cc}$ is increased. The variance of $M_{cc}$
then decreases when we increase $D$ further, because the value of
$M_{cc}$ is too small in this case and the noise-induced
fluctuations in $M_{cc}$ are suppressed.

To give more insights for synchronization with optically stochastic noises
in our $\mathcal{PT}$-symmetric optomechnical system, we show
additional analysis of another index of synchronization---the
Kramers rate, which is more suitable to describe noisy
synchronized systems.
The Kramers rates of two subsystems are alternative indices to
show the correlation between two subsystems. When the Kramers
rates of two subsystems coincide with each other, the two
subsystems are well correlated~\cite{Neiman1}. We then calculate
the Kramers rates $r_1$ and $r_2$ of the mechanical displacements
$x_1$ and $x_2$, respectively. The Kramers rate is originally
defined as the transition rate between neighboring potential wells
of a particle caused by stochastic forces, which was first
proposed by Kramers in 1940~\cite{Kramers}.

\begin{figure}[h]
\centerline{
\includegraphics[width=8.6 cm,clip]{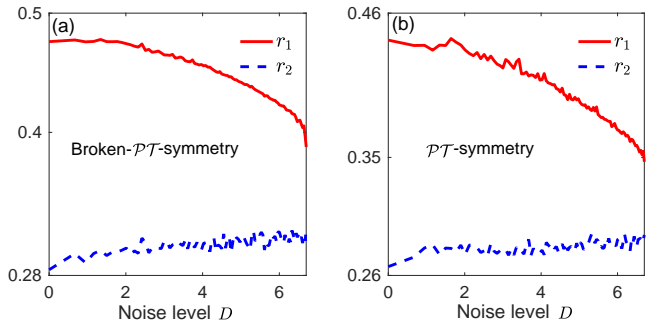}}
\caption{(Color online) The Kramers rates $r_1$ and $r_2$ of
mechanical displacements $x_1$ and $x_2$ versus the noise
intensity $D$ in broken-$\mathcal{PT}$-symmetric and
$\mathcal{PT}$-symmetric regime. (a) The red solid curve (blue
dashed curve) represents the curve for Kramers rate $r_1$ ($r_2$)
versus the noise intensity $D$ in the
broken-$\mathcal{PT}$-symmetric regime. Here the optical coupling
strength $\kappa=27.76$ MHz is fixed. (b) The  Kramers rates $r_1$ and $r_2$
with different stochastic noise intensity $D$ correspond to the
$\mathcal{PT}$-symmetric regime, where the optical coupling
strength is fixed as $\kappa=32.19$ MHz.}\label{Fig6}
\end{figure}

Here, we use the mean first passage time \cite{Klein,Hofmann},
i.e., the average time that the particle moves from one potential
well to the other well, to evaluate the Kramers rates $r_1$ and
$r_2$ of mechanical displacements $x_1$ and $x_2$. We obtain the
histograms of $x_{1,2}$ through numerical simulation first, and
then find out the locations of with
the maximum probability of $x_{1,2}$, i.e., the potential wells of
$x_{1,2}$, based on the distribution of histograms, by which we
can obtain the mean first passage times $\tau_{1,2}$, i.e., the
average value of the time intervals between two potential
wells for each mechanical displacement.
The Kramers rates $r_1$ and $r_2$ can then be calculated by the
reciprocal of the mean first passage times $\tau_{1,2}$, i.e.,
$r_{i}=1/\tau_i$ ($i=1,2$). The simulation results for $r_1$ and
$r_2$ are presented in Fig.~\ref{Fig6}.  It can be seen that,
both in broken-$\mathcal{PT}$-symmetric ( Fig.~\ref{Fig6}(a))
and $\mathcal{PT}$-symmetric ( Fig.~\ref{Fig6}(b)) regimes,
the Kramers rates $r_1$ and $r_2$ get closer with the increase of the
noise intensity, which means that the partial frequencies of the
mechanical displacements $x_1$ and $x_2$ get closer when the noise
intensity $D$ is increased. It means that the optically stochastic noises can
improve the correlation between $x_1$ and $x_2$.

\subsection{Thermal noises in the mechanical modes}

In the above analysis we do not consider the effects of the thermal noises
in the mechanical modes. Actually, these thermal noises in the mechanical
modes can also benefit the synchronization between the two mechanical
modes in our $\mathcal{PT}$-symmetric optomechanical system.
In order to simplify our discussions, we only consider the thermal noises
in the mechanical modes in this section, and assume that the thermal noises in the
mechanical modes are white noises, based on which the Langevin equation of the
mechanical modes can be expressed as
\begin{eqnarray}\label{The reduced Langevin equation with Brownian noises}
\ddot{x}_1&=&-2\Gamma_{m}\dot{x}_1-\tilde{\Omega}_1^2x_1-\kappa_{\rm{mech}} x_2 +\Gamma_{\rm{noise}1}(t),\nonumber\\
\ddot{x}_2&=&-2\Gamma_{m}\dot{x}_2-\tilde{\Omega}_2^2x_2-\kappa_{\rm{mech}} x_1 +\Gamma_{\rm{noise}2}(t),
\end{eqnarray}
where the constant driving terms induced by optical modes have been included
into $x_{1,2}$ by a coordinate transformation for simplicity.
The mechanical damping rate $\Gamma_{m}$ includes the damping rate shift $\delta\Gamma_m$
induced by the corresponding thermal noise, i.e., $\Gamma_m=\Gamma_{mo}+\delta\Gamma_m$,
where $\Gamma_{mo}$ is the original mechanical damping rate without considering thermal noise.
The mechanical thermal noises $\Gamma_{\rm{noise1}}$ and $\Gamma_{\rm{noise2}}$ are
diffusion terms with $\delta$-correlated Gaussian distribution
\begin{eqnarray}
\left\langle \Gamma_{\rm{noise}\ i}(t)  \right\rangle &=& 0,\nonumber\\
\left\langle \Gamma_{\rm{noise}\ i}(t)\Gamma_{\rm{noise}\ j}(t')  \right\rangle &=& 4\Gamma_m k T\delta(t-t'),
\end{eqnarray}
where $k$ is the Boltzman's constant and $T$ is the temperature.

\begin{figure}[h]
\centerline{
\includegraphics[width=8.6 cm,clip]{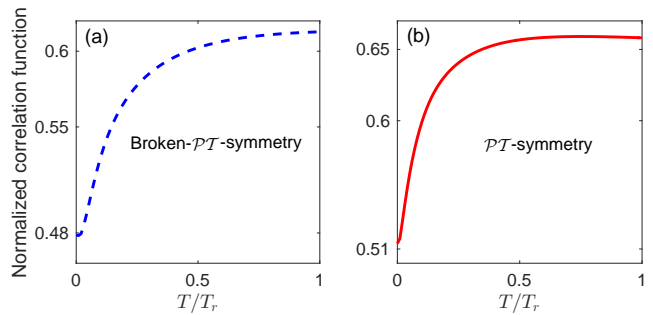}}
\caption{(Color online) Numerical results of
the normalized correlation function $R$ with different values
of temperature $T$ in broken-$\mathcal{PT}$-symmetric and
$\mathcal{PT}$-symmtric regimes, where $T_r$ denotes the room temperature.
(a) Effects of the thermal noises on $R$ with respect to different
temperature $T$ in broken-$\mathcal{PT}$-symmetric regime
with $\kappa=27.76$ MHz.
(b) Effects of the thermal noises on $R$ versus different $T$
in $\mathcal{PT}$-symmetric regime with $\kappa=32.19$ MHz.
}  \label{Fig7}
\end{figure}

To show the positive influence of thermal noises on the synchronization,
we present the numerical results of the normalized correlation function $R$~\cite{Risken}
between the two mechanical oscillators in Figs.~\ref{Fig7}(a) and (b)
by changing the temperature $T$ and fixing other parameters in both
broken-$\mathcal{PT}$-symmetric and $\mathcal{PT}$-symmetric regimes, where
$T_r$ is the room temperature. In the broken-$\mathcal{PT}$-symmetric regime
with optical coupling strength $\kappa=27.76$ MHz, $R$ (blue-dashed curve)
is enhanced with increasing temperature $T$, and reaches 0.61 at the
room temperature $T_r$, which is larger than 0.48 when we ignore the thermal
noises. Similarly, in the $\mathcal{PT}$-symmetric regime with optical
coupling strength $\kappa=32.19$ MHz, $R$ (red-solid curve) increases with
temperature $T$, and reaches 0.65 at the room temperature,
which is larger than 0.51 when we ignore the thermal noises. It means that
the thermal noises in the mechanical modes can also benefit the synchronization
between the two mechanical modes in our optomechanical $\mathcal{PT}$-symmetric system.

\begin{figure}[h]
\centerline{
\includegraphics[width=8.6 cm,clip]{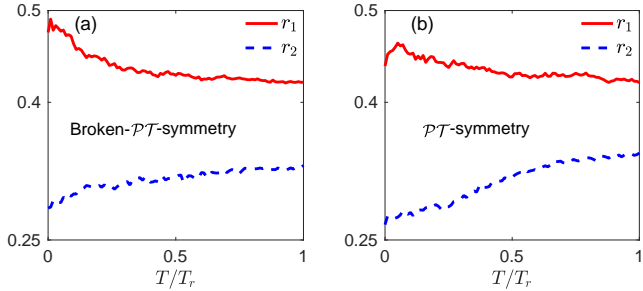}}
\caption{(Color online) The Kramers rates $r_1$ and $r_2$ of mechanical
displacements $x_1$ and $x_2$ versus the temperature $T$ in both broken-$\mathcal{PT}$-symmetric
and $\mathcal{PT}$-symmetric regime, where $T_r$ is the room temperature.
(a) The red solid curve (blue dashed curve) denotes the Kramers rate $r_1$ ($r_2$) with
increasing the temperature $T$ in the broken-$\mathcal{PT}$-symmetric regime,
where the optical coupling strength $\kappa=27.76$ MHz is fixed.
(b) The Kramers rates $r_1$ and $r_2$ versus the temperature $T$ correspond to the
$\mathcal{PT}$-symmetric regime ($\kappa=32.19$ MHz).}\label{Fig8}
\end{figure}

To give more insights into the phenomenon presented,
we calculate the Kramers rates $r_1$ and $r_2$ of mechanical displacements
$x_1$ and $x_2$. The simulation results
for Kramers rates $r_1$ and $r_2$ are shown in Figs.~\ref{Fig8}(a) and (b).
In Fig.~\ref{Fig8}(a), the red solid curve denotes Kramers rate $r_1$
with different values of temperature $T$ in the broken-$\mathcal{PT}$-symmetric
regime with optical coupling strength $\kappa=27.76$ MHz,
and the blue dashed curve corresponds to the Kramers rate $r_2$.
We can see in Fig.~\ref{Fig8}(a) that Kramers rates $r_1$ and $r_2$
tend to get closer to each other as the temperature $T$ increases to the
room temperature $T_r$.
Similar phenomenon can be observed in the $\mathcal{PT}$-symmetric regime
as shown in Fig.~\ref{Fig8}(b), i.e., the mechanical thermal noises
tend to decrease the difference between the Kramers rates $r_1$ and $r_2$
as the temperature increases to the room temperature, where the optical
coupling strength is fixed as $\kappa=32.19$ MHz.
These simulation results indicate that more mechanical thermal noises
can lead the partial frequencies of the two mechanical displacements
$x_1$ and $x_2$ to tend to  be consistent with each other, and thus
benefit the synchronization in our $\mathcal{PT}$-symmetric optommechanical system.

Furthermore, we can also observe the beneficial effect
of the mechanical thermal noises on the synchronization by
theoretically analyzing the correlation function between the
two mechanical modes, when we consider small time $t$.
Actually, at small time limit \cite{Risken}, the normalized
correlation function between the two mechanical modes can be
approximated as (see the derivations in Appendix \ref{Derivation of the normalized correlation function})
\begin{eqnarray}\label{Normalized correlation function}
R(\tau,t)&\approx& 1-2\tilde\Omega_1^2\tau t + \frac{q}{2} \kappa_{\rm{mech}}\tilde\Omega_1^2\tau t^2    +\frac{q}{3} \kappa_{\rm{mech}}\tilde\Omega_1^2 \tau t^3\nonumber\\
&=&1-2\tilde\Omega_1^2\tau t+2\Gamma_m k T\kappa_{\rm{mech}}\tilde\Omega_1^2\tau t^2\nonumber\\
&&+\frac{4}{3}\Gamma_m k T\kappa_{\rm{mech}}\tilde\Omega_1^2 \tau t^3,
\end{eqnarray}
where $q$ is the intensity of the mechanical thermal noises, i.e., $q=4\Gamma_m k T$.
It is shown in Eq.~(\ref{Normalized correlation function}) that
the normalized correlation function $R$ can be enhanced by the
increase of the intensity of the thermal noises, which is
in consonance with the above simulation results, as shown
in Figs.~\ref{Fig7} and \ref{Fig8}. It proves that the thermal
noises in the mechanical modes can benefit the synchronization
in our $\mathcal{PT}$-symmetric synchronization system.

\section{Conclusion and discussion}\label{s5}
We have shown that the mechanical motions of
two coupled $\mathcal{PT}$-symmetric optomechanical resonators
with far-off-resonant mechanical frequencies can be synchronized
when the system approaches the EP. In particular, in the
$\mathcal{PT}$-symmetric regime, the two mechanical modes are
easier to be synchronized with weaker optical coupling strength
between the two optomechanical resonators. Additionally, it is
shown that noises will be enhanced in the vicinity of the EP in our
system, and the enhanced noises will benefit the synchronization
process if only the strengths of the noises are not too strong.
Our study opens up a new dimension of research for
$\mathcal{PT}$-symmetric optomechanical system for possible
applications such as metrology, cooling, and communication.
It also gives new perspectives for synchronization in optomechanical systems.

\section{Acknowledgments}\label{s6}
JZ is supported by the NSFC under Grant Nos. 61622306, 11674194.
YXL and JZ are supported by the National Basic Research Program of China (973
Program) under Grant No. 2014CB921401, the Tsinghua University
Initiative Scientific Research Program, and the Tsinghua National
Laboratory for Information Science and Technology (TNList)
Cross-discipline Foundation. JZ is also supported by the Youth
Innovation Fund of Beijing National Research Center for
Information Science and Technology (BNRist). LY is supported by
the NSF grant No. EFMA1641109, ARO grant No. W911NF1210026 and ARO
grant No. W911NF1710189.

\appendix

\section{Weaker condition of $\mathcal{PT}$-symmetry with $\Delta_1 \ne \Delta_2$}\label{Weaker condition of PT-symmetry}
Generally, in our optomechanical system if we consider symmetric
optical driving detunings $\Delta_1=\Delta_2$, there exists an exceptional
point where the two optical supermodes degenerate with each other at this
point. However, if $\Delta_{-}=|\Delta_2-\Delta_1|\ne 0$, the degeneracy
of the optical supermodes at the previous exceptional point will be broken.
Now we prove that even though the two optical driving detunings are asymmetric, i.e.,
$\Delta_{-} \ne 0$, the non-degeneracy can be small enough that the characteristic
of $\mathcal{PT}$-symmetry can still be maintained under a weaker condition,
i.e., $\Delta_{-}$ is small enough.

In order to analyze the $\mathcal{PT}$-symmetric structure in our optomechanical
system, we consider the optical modes only and assume that the nonlinear optomechanical
interaction between the optical mode and mechanical mode is a nonlinearly
induced frequency shift for the corresponding optical mode in each cavity~\cite{Schonleber}.
Here, we treat the stationary state $\beta_{1s}$ ($\beta_{2s}$) of the mechanical
mode $\beta_{1}$ ($\beta_{2}$) as a parameter which leads to a frequency detuning
$\Delta_{1s}=g_{om}(\beta_{1s}+\beta_{1s}^{*})$ ($\Delta_{2s}=g_{om}(\beta_{2s}+\beta_{2s}^{*})$)
for the optical mode $\alpha_{1}$ ($\alpha_{2}$).
By taking $\dot{\alpha}_{1,2}=\dot{\beta}_{1,2}=0$ in Eqs.~\ref{Dynamical Equations of the optomechanical system},
we can obtain the stationary states of the optical and mechanical modes
which satisfies the following equations
\begin{eqnarray}\label{The complete stationary equations}
0&=&(\gamma_1-i\Delta_1)\alpha_{1s}-i\kappa \alpha_{2s}-i g_{om}\alpha_{1s}(\beta_{1s}+\beta_{1s}^{*}) \nonumber\\
&&+\sqrt{2\gamma_{1ex}}\epsilon_1,\nonumber\\
0&=&-(\gamma_2+i\Delta_2)\alpha_{2s}-i\kappa \alpha_{1s}-i g_{om} \alpha_{2s}(\beta_{2s}+\beta_{2s}^{*})\nonumber\\
&&+\sqrt{2\gamma_{2ex}\epsilon_2}, \nonumber\\
0&=&-(\Gamma_{m1}+i\Omega_1)\beta_{1s}-i g_{om}|\alpha_{1s}|^2, \nonumber\\
0&=&-(\Gamma_{m2}+i\Omega_2)\beta_{2s}-i g_{om}|\alpha_{2s}|^2.
\end{eqnarray}
By solving the above the equation, the stationary states of the
mechanical modes can be expressed as
\begin{eqnarray}
\beta_{1s}&=&-g_{om}\frac{\Omega_1+i\Gamma_{m1}}{\Gamma_{m1}^{2}+\Omega_{1}^{2}}\left|\alpha_{1s} \right|^2, \nonumber \\
\beta_{2s}&=&-g_{om}\frac{\Omega_2+i\Gamma_{m2}}{\Gamma_{m2}^{2}+\Omega_{2}^{2}}\left|\alpha_{2s} \right|^2,
\end{eqnarray}
and the stationary states of the optical modes satisfy the following equations
\begin{eqnarray}\label{Requirements of the stationary states of optical modes}
\left[ \gamma_1-i(\Delta_1+\Delta_{1s}) \right]\alpha_{1s} - i\kappa\alpha_{2s} +\sqrt{2\gamma_{1ex}}\epsilon_1&=&0, \nonumber\\
\left[-\gamma_2-i(\Delta_2+\Delta_{2s}) \right]\alpha_{2s} - i\kappa\alpha_{1s} +\sqrt{2\gamma_{2ex}}\epsilon_2&=&0, \nonumber\\
\end{eqnarray}
where
\begin{eqnarray}
\Delta_{1s}&=&-\frac{2\Omega_1g_{om}^2}{\Gamma_{m1}^2+\Omega_1^2}\left| \alpha_{1s} \right|^2, \nonumber \\
\Delta_{2s}&=&-\frac{2\Omega_2g_{om}^2}{\Gamma_{m2}^2+\Omega_2^2}\left| \alpha_{2s} \right|^2.
\end{eqnarray}
By substituting the stationary states $\beta_{1s}$ and $\beta_{2s}$
into Eq.~(\ref{Dynamical Equations of the optomechanical system}) and
eliminating the mechanical modes, we have
\begin{eqnarray}\label{Purely_optical_structure}
\dot{\alpha}_1&=&\left[ \gamma_1-i(\Delta_1+\Delta_{1s}) \right]\alpha_1
-i\kappa \alpha_2+\sqrt{2\gamma_{1ex}}\epsilon_1,  \nonumber \\
\dot{\alpha}_2&=&\left[ -\gamma_2-i(\Delta_2+\Delta_{2s}) \right]\alpha_2
-i\kappa \alpha_1+\sqrt{2\gamma_{2ex}}\epsilon_2. \nonumber \\
\end{eqnarray}

Based on Eq.~(\ref{Purely_optical_structure}),
we can calculate the eigenfrequencies of the optical supermodes as
\begin{align}\label{Eigenfrequencies of the optical supermodes}
& \omega_{o+}=\frac{\gamma_1-\gamma_2}{2}-i\frac{\Delta_1+\Delta_{1s}+\Delta_2+\Delta_{2s}}{2} \nonumber\\
&\hspace{3mm} +\sqrt{\left[\frac{\gamma_1+\gamma_2}{2}+i\left(\frac{\Delta_2+\Delta_{2s}}{2} -\frac{\Delta_1+\Delta_{1s}}{2} \right) \right]^2-\kappa^2},\nonumber\\
& \omega_{o-}=\frac{\gamma_1-\gamma_2}{2}-i\frac{\Delta_1+\Delta_{1s}+\Delta_2+\Delta_{2s}}{2} \nonumber\\
&\hspace{3mm} -\sqrt{\left[\frac{\gamma_1+\gamma_2}{2}+i\left(\frac{\Delta_2+\Delta_{2s}}{2}
-\frac{\Delta_1+\Delta_{1s}}{2} \right) \right]^2-\kappa^2}.
\end{align}
Considering the balanced gain and loss ($\gamma_1=\gamma_2=\gamma$),
the above equations can be reduced to
\begin{align}\label{Eigenfrequencies of the optical supermodes with balanced gain and loss}
& \omega_{o+}=-i\frac{\Delta_1+\Delta_{1s}+\Delta_2+\Delta_{2s}}{2} \nonumber\\
&\hspace{3mm} +\sqrt{\left[\gamma+i\left(\frac{\Delta_2+\Delta_{2s}}{2} -\frac{\Delta_1+\Delta_{1s}}{2} \right) \right]^2-\kappa^2},\nonumber\\
& \omega_{o-}=-i\frac{\Delta_1+\Delta_{1s}+\Delta_2+\Delta_{2s}}{2} \nonumber\\
&\hspace{3mm} -\sqrt{\left[\gamma+i\left(\frac{\Delta_2+\Delta_{2s}}{2}
-\frac{\Delta_1+\Delta_{1s}}{2} \right) \right]^2-\kappa^2}.
\end{align}
Actually, the vacuum optomechanical coupling $g_{om}$ in general optical
cavities~\cite{Aspelmeyer} is very small, thus if $\Delta_{-}=\Delta_2-\Delta_1$
is small enough, the imaginary part in the root sign in Eq.~\ref{Eigenfrequencies of the optical supermodes with balanced gain and loss}
can be ignored, and the eigenvalues can be reduced to
\begin{align}\label{Eigenfrequencies of the optical supermodes with balanced gain and loss2}
& \omega_{o+}\approx -i\frac{\Delta_1+\Delta_2}{2}+\sqrt{\gamma^2-\kappa^2} \nonumber\\
& \omega_{o-}\approx -i\frac{\Delta_1+\Delta_2}{2}-\sqrt{\gamma^2-\kappa^2}.
\end{align}
It means that the two eigenvalues of optical supermodes tend to degenerate with each other
at the exceptional point $\kappa=\gamma$. In fact, by substituting $\alpha_{1s}$ and
$\alpha_{2s}$ into Eqs.~\ref{Eigenfrequencies of the optical supermodes with balanced gain and loss},
the non-degeneracy of the optical supermodes at the exceptional point ($\kappa=\gamma$)
can be evaluated as
\begin{equation}
\frac{\Delta_{\rm{split}}}{\gamma}\approx \sqrt{\frac{\Delta_{-}^3}{\frac{2}{3}\gamma
\left( g_{om}^2\frac{\Omega_2+\Omega_1}{\Omega_1\Omega_2} \gamma\epsilon^2\right)^{2/3}}},
\end{equation}
thus when
\begin{equation}\label{Weaker condition 2}
g_{om} \ll \Delta_{-}\ll \sqrt[3]{\frac{2}{3}\gamma\left(g_{om}^{2}\frac{\Omega_2+\Omega_1}{\Omega_1\Omega_2}\gamma\epsilon^2 \right)^{2}} \ll \gamma,
\end{equation}
where $\Delta_{\rm{split}}=\rm{Im}[\omega_{o+}-\omega_{0-}]=\rm{Re}[\omega_{o+}-\omega_{0-}]$.
It can be inferred that this non-degeneracy can be very small that
the properties of $\mathcal{PT}$-symmetric structure can be greatly held in our optomechanical system.
We name the condition of Eq.~\ref{Weaker condition 2} as weaker condition for $\mathcal{PT}$-symmetry
in our optomechanical system, and it can be easily realized in general cavity optomechanical systems.

As for the simulation results in Figs.~\ref{Fig2} (a) and (b) in the main text, we first calculate
the stationary states of $\alpha_{1s}$ and $\alpha_{2s}$ by numerically solving
the Eqs.~\ref{Requirements of the stationary states of optical modes}, then
the eigenvalues of optical supermodes can be obtained by substituting
$\alpha_{1s}$ and $\alpha_{2s}$ into Eqs.~\ref{Eigenfrequencies of the optical supermodes with balanced gain and loss}.

\section{The derivation of the reduced dynamical equation of the mechanical modes}\label{The derivation of the reduced dynamical equation of the mechanical modes}
Based on the dynamical equation in Eq.~(\ref{Dynamical Equations of the optomechanical system}),
we can adiabatically eliminate the degrees of freedom of the optical
modes, and derive the reduced dynamical equations of the mechanical modes.
In fact, by rewriting the first two equations in
Eq.~(\ref{Dynamical Equations of the optomechanical system}) in matrix format, we have
\begin{eqnarray}\label{Matrix form of movement equations about optical modes}
\left[\begin{array}{cc}
\dot \alpha_1\\
\dot \alpha_2
\end{array}\right]&=&M
\left[\begin{array}{cc}
\alpha_1\\
\alpha_2
\end{array}\right]
+\left[\begin{array}{cc}
-i g_{om}\alpha_1(\beta_1+\beta_{1}^*)\nonumber\\
-i g_{om}\alpha_2(\beta_2+\beta_{2}^*)
\end{array}\right] \\
&&+\left[\begin{array}{cc}
\sqrt{2\gamma_{1ex}}\epsilon_1\\
\sqrt{2\gamma_{2ex}}\epsilon_2
\end{array}\right],
\end{eqnarray}
where
\begin{eqnarray}
M=\left[\begin{array}{cc}
\gamma_1-i\Delta_1 & -i\kappa\\
-i\kappa & -\gamma_2-i\Delta_2
\end{array}\right]. \nonumber
\end{eqnarray}
The matrix $M$ can be diagonalized as
\begin{eqnarray}
M=T\Lambda T^{-1}, \nonumber
\end{eqnarray}
where
\begin{eqnarray}
\Lambda=\left[\begin{array}{cc}
\omega_{+} & 0\\
0 & \omega_{-}
\end{array}\right],
T=\left[\begin{array}{cc}
\tau_{+} & \tau_{-}\\
1 & 1
\end{array}\right],  \nonumber
\end{eqnarray}
and
\begin{eqnarray}
\omega_{+}&=&\frac{\gamma_1-\gamma_2}{2}-i\frac{\Delta_1+\Delta_2}{2} \nonumber\\
&&-i\sqrt{\kappa^2+\left(\frac{\Delta_1-\Delta_2}{2}+i\frac{\gamma_1+\gamma_2}{2}\right)^2}, \nonumber\\
\omega_{-}&=&\frac{\gamma_1-\gamma_2}{2}-i\frac{\Delta_1+\Delta_2}{2} \nonumber\\
&&+i\sqrt{\kappa^2+\left(\frac{\Delta_1-\Delta_2}{2}+i\frac{\gamma_1+\gamma_2}{2}\right)^2}, \nonumber\\
\tau_{+}&=&\frac{\Delta_1-\Delta_2+i(\gamma_1+\gamma_2)}{2\kappa} \nonumber\\
&&+\sqrt{1+\left(\frac{\Delta_1-\Delta_2+i(\gamma_1+\gamma_2)}{2\kappa}\right)^2},  \nonumber\\
\tau_{-}&=&\frac{\Delta_1-\Delta_2+i(\gamma_1+\gamma_2)}{2\kappa} \nonumber\\
&&-\sqrt{1+\left( \frac{\Delta_1-\Delta_2+i(\gamma_1+\gamma_2)}{2\kappa} \right)^2}.  \nonumber
\end{eqnarray}
Thus, we can introduce the following optical supermodes
\begin{eqnarray}\label{transformation between normal modes and supermodes}
\left[\begin{array}{cc}
\alpha_{+}\\
\alpha_{-}
\end{array}\right]=
T^{-1}\left[\begin{array}{cc}
\alpha_1\\
\alpha_2
\end{array}\right],
\end{eqnarray}
by which Eq.~(\ref{Matrix form of movement equations about optical
modes}) can be reexpressed as
\begin{align*}
&\left[\begin{array}{cc}
\dot{\alpha}_{+}\\
\dot{\alpha}_{-}
\end{array}\right]=
\left[\begin{array}{cc}
\omega_{+} & 0 \\
0 & \omega_{-}
\end{array}\right]
\left[\begin{array}{cc}
\alpha_{+}\\
\alpha_{-}
\end{array}\right] -i g_{om}\times\\
&\hspace{2mm} \left[\begin{array}{cc}
(\lambda_{+}\alpha_{+}+\lambda_{-} \alpha_{-})(\beta_1+\beta_1^{*})-\lambda_{-}(\alpha_{+}+\alpha_{-})(\beta_2+\beta_2^{*})\\
-(\lambda_{+}\alpha_{+}+\lambda_{-}\alpha_{-})(\beta_1+\beta_1^{*})+\lambda_{+}(\alpha_{+}+\alpha_{-})(\beta_2+\beta_2^{*})
\end{array}\right]   \nonumber\\
&\hspace{2mm} +\left[\begin{array}{cc}
\mu\sqrt{2\gamma_{1ex}}\epsilon_1-\lambda_{-}\sqrt{2\gamma_{2ex}}\epsilon_2\\
-\mu\sqrt{2\gamma_{1ex}}\epsilon_1+\lambda_{+}\sqrt{2\gamma_{2ex}}\epsilon_2
\end{array}\right], \nonumber
\end{align*}
where
\begin{eqnarray}
\lambda_{-}&=&\frac{\Delta_1-\Delta_2+i(\gamma_1+\gamma_2)-\Xi_1}{2\Xi_1},   \nonumber\\
\lambda_{+}&=&\frac{\Delta_1-\Delta_2+i(\gamma_1+\gamma_2)+\Xi_1}{2\Xi_1},  \nonumber\\
\mu&=&\frac{\kappa}{\Xi_1},   \nonumber
\end{eqnarray}
where
\begin{equation}
\Xi_1=\sqrt{4\kappa^2+\left(\Delta_1-\Delta_2+i(\gamma_1+\gamma_2) \right)^2}.\nonumber
\end{equation}

To adiabatically eliminate the degrees of freedom of the optical modes,
we let $\dot{\alpha}_{+}=\dot{\alpha}_{-}=0$, by which we can obtain the
following stationary solution
\begin{eqnarray}
\alpha_{+ss}&=&\frac{-\mu\left(\omega_{-}-i g_{om}(\beta_2+\beta_2^{*}) \right)\sqrt{2\gamma_{1ex}}\epsilon_1}{\Xi_2}   \nonumber\\
&&+\frac{\lambda_{-}\left(\omega_{-}-i g_{om}(\beta_1+\beta_1^{*}) \right)\sqrt{2\gamma_{2ex}}\epsilon_2}{\Xi_2},\nonumber\\
\alpha_{-ss}&=&\frac{\mu\left(\omega_{+}-i g_{om}(\beta_2+\beta_2^{*})\right)\sqrt{2\gamma_{1ex}}\epsilon_1}{\Xi_2}\nonumber \\
&&-\frac{\lambda_{+}\left(\omega_{+}-i
g_{om}(\beta_1+\beta_1^{*}) \right)\sqrt{2\gamma_{2ex}}\epsilon_2}{\Xi_2}
\end{eqnarray}
where
\begin{eqnarray}
\Xi_2&=&\omega_{+}\omega_{-}+i g_{om}(\omega_{+}\lambda_{-}-\omega_{-}\lambda_{+})(\beta_1+\beta_1^{*}) \nonumber\\
&&+i g_{om}(-\omega_{+}\lambda_{+}+\omega_{-}\lambda_{+})(\beta_2+\beta_2^{*})\nonumber\\
&&-g_{om}^{2}(\lambda_{+}-\lambda_{-})^2(\beta_1+\beta_1^{*})
(\beta_2+\beta_2^{*})
\end{eqnarray}
By introducing the power-series expansion and omitting high-order
terms of $\beta_1$ and $\beta_2$ ($g_{om}\ll |\Delta_2-\Delta_1|$), the above solutions can be
simplified as
\begin{eqnarray}
\alpha_{+ss}&\approx& -\frac{\mu\sqrt{2\gamma_{1ex}}\epsilon_1-\lambda_{-}\sqrt{2\gamma_{2ex}}\epsilon_2}{\omega_{+}} \nonumber\\
&+&i g_{om}\frac{\Xi_3-\lambda_{-}\omega_{+}\omega_{-}\sqrt{2\gamma_{2ex}}\epsilon_2}{\left(\omega_{+}\omega_{-} \right)^2}(\beta_1+\beta_1^{*})  \nonumber\\
&+&i g_{om}\frac{-\Xi_3+\mu\omega_{+}\omega_{-}\sqrt{2\gamma_{1ex}}\epsilon_1}
{\left(\omega_{+}\omega_{-} \right)^2}(\beta_2+\beta_2^{*}),   \nonumber\\
\alpha_{-ss}&\approx& -\frac{\mu\sqrt{2\gamma_{1ex}}\epsilon_1-\lambda_{+}\sqrt{2\gamma_{2ex}}\epsilon_2}{\omega_{-}} \nonumber\\
&+&i g_{om}\frac{\Xi_4-\lambda_{+}\omega_{+}\omega_{-}\sqrt{2\gamma_{2ex}}\epsilon_2}
{\left(\omega_{+}\omega_{-}\right)^2}(\beta_1+\beta_1^{*})    \nonumber\\
&+&i g_{om}\frac{-\Xi_4
+\mu\omega_{+}\omega_{-}\sqrt{2\gamma_{1ex}}\epsilon_1}{\left(\omega_{+}\omega_{-} \right)^2}(\beta_2+\beta_2^{*}),
\end{eqnarray}
where
\begin{eqnarray}
\Xi_3&=&\omega_{-}\left(\mu\sqrt{2\gamma_{1ex}}\epsilon_1-\lambda_{-}\sqrt{2\gamma_{2ex}}\epsilon_2 \right)
\left(\omega_{+}\lambda_{-}-\omega_{-}\lambda_{+} \right),\nonumber\\
\Xi_4&=&\omega_{+}(\mu\sqrt{2\gamma_{1ex}}\epsilon_1-\lambda_{+}\sqrt{2\gamma_{2ex}}\epsilon_2)
(\omega_{+}\lambda_{-}-\omega_{-}\lambda_{+}).\nonumber
\end{eqnarray}
Thus, the stationary solutions of $\alpha_{1}$ and $\alpha_{2}$
can be expressed as
\begin{eqnarray}
\alpha_{1ss}&=&\tau_{+}\alpha_{+ss}+\tau_{-}\alpha_{-ss}    \nonumber\\
&=&\frac{\sigma_2\sqrt{2\gamma_{1ex}}\epsilon_1-i\kappa\sqrt{2\gamma_{2ex}}\epsilon_2}{\kappa^2+\delta^2+\sigma^2}\nonumber\\
&&-i g_{om}\frac{\sigma_{2}^2\sqrt{2\gamma_{1ex}}\epsilon_1-i\kappa\sigma_2\sqrt{2\gamma_{2ex}}\epsilon_2}
{\kappa^2+\delta^2+\sigma^2}(\beta_1+\beta_1^{*}) \nonumber\\
&&+i g_{om}\frac{\kappa^2\sqrt{2\gamma_{1ex}}\epsilon_1+i\kappa\sigma_1\sqrt{2\gamma_{2ex}}\epsilon_2}
{\kappa^2+\delta^2+\sigma^2}(\beta_2+\beta_2^{*}), \nonumber\\
\alpha_{2ss}&=&\alpha_{+ss}+\alpha_{-ss}   \nonumber\\
&=&\frac{-i\kappa\sqrt{2\gamma_{1ex}}\epsilon_1+\sigma_1\sqrt{2\gamma_{2ex}}\epsilon_2}{\kappa^2+\delta^2+\sigma^2}\nonumber\\
&&+i g_{om}\frac{i\kappa\sigma_2\sqrt{2\gamma_{1ex}}\epsilon_1+\kappa^2\sqrt{2\gamma_{2ex}}\epsilon_2}
{\kappa^2+\delta^2+\sigma^2}(\beta_1+\beta_1^{*}) \nonumber\\
&&-i g_{om}\frac{-i\kappa\sigma_1\sqrt{2\gamma_{1ex}}\epsilon_1+\sigma_1^2\sqrt{2\gamma_{2ex}}\epsilon_2}
{\kappa^2+\delta^2+\sigma^2}(\beta_2+\beta_2^{*}), \nonumber\\
\end{eqnarray}
where
\begin{eqnarray}
\sigma_1=-\gamma_1+i\Delta_1, && \sigma_2=\gamma_2+i\Delta_2,  \nonumber\\
\delta=\frac{\Delta_1-\Delta_2}{2}+i\frac{\gamma_1+\gamma_2}{2},
&&
\sigma=\frac{-\gamma_1+\gamma_2}{2}+i\frac{\Delta_1+\Delta_2}{2}.
\nonumber
\end{eqnarray}
By substituting the above stationary solution into the dynamical
equations of the mechanical modes $\beta_1$ and $\beta_2$ in
Eq.~(\ref{Dynamical Equations of the optomechanical system}),
and dropping the counter-rotating terms with $\beta_{1,2}^{*}$,
the dynamical equation of reduced mechanical system
can be expressed in the matrix format as
\begin{eqnarray}\label{Reduced mechanical motion equations}
\left[\begin{array}{cc}
\dot \beta_1\\
\dot \beta_2
\end{array}\right]&=&
\left[\begin{array}{cc}
-\Gamma_{m1}-i(\Omega_1+\delta\Omega_1) & \kappa_{\rm{mech}} \\
\kappa_{\rm{mech}} & -\Gamma_{m2}-i(\Omega_2+\delta\Omega_2)
\end{array} \right] \nonumber\\
&&\times\left[\begin{array}{cc}
\beta_1\\
\beta_2
\end{array}\right]-
\left[\begin{array}{cc}
i\eta_1\\
i\eta_2
\end{array}\right],
\end{eqnarray}
where
\begin{align}\label{Parameters of reduced mechanical motion equations}
&\delta\Omega_1=\frac{4g_{om}^2\left[(\kappa^2-\Delta_1\Delta_2)\Delta_2-\Delta_1\gamma_2^2\right]}
{\left[(\kappa^2-\Delta_1\Delta_2-\gamma_1\gamma_2)^2+(\Delta_1\gamma_2-\Delta_2\gamma_1)^2\right]^2},   \nonumber\\
&\hspace{4mm}\times\left[\gamma_2^2\gamma_{1ex}\epsilon_1^2+(\Delta_2\sqrt{\gamma_{1ex}}\epsilon_1-\kappa\sqrt{\gamma_{2ex}}\epsilon_2)^2\right], \nonumber\\
&\delta\Omega_2=\frac{4g_{om}^2\left[(\kappa^2-\Delta_1\Delta_2)\Delta_1-\Delta_2\gamma_1^2\right]}
{\left[(\kappa^2-\Delta_1\Delta_2-\gamma_1\gamma_2)^2+(\Delta_1\gamma_2-\Delta_2\gamma_1)^2\right]^2},  \nonumber\\
&\hspace{4mm}\times \left[\gamma_1^2\gamma_{2ex}\epsilon_2^2+(\kappa\sqrt{\gamma_{1ex}}\epsilon_1-\Delta_1\sqrt{\gamma_{2ex}}\epsilon_2)^2\right], \nonumber\\
&\kappa_{\rm{mech}} =4g_{om}^2\kappa\times \nonumber\\ &\hspace{4mm}\bigg[\frac{\kappa\gamma_{1ex}\epsilon_1^2\left[\Delta_2(\kappa^2-\Delta_1\Delta_2-2\gamma_1\gamma_2)+\Delta_1\gamma_2^2\right]}
{\left[(\kappa^2-\Delta_1\Delta_2-\gamma_1\gamma_2)^2+(\Delta_1\gamma_2-\Delta_2\gamma_1)^2\right]^2}  \nonumber\\
&\hspace{4mm}+\frac{+\kappa\gamma_{2ex}\epsilon_2^2\left[\Delta_1(\kappa^2-\Delta_1\Delta_2)-\Delta_2\gamma_1^2\right]}
{\left[(\kappa^2-\Delta_1\Delta_2-\gamma_1\gamma_2)^2+(\Delta_1\gamma_2-\Delta_2\gamma_1)^2\right]^2} \nonumber\\
&\hspace{4mm}-\frac{\sqrt{\gamma_{1ex}\gamma_{2ex}}
\epsilon_1\epsilon_2}
{\left[(\kappa^2-\Delta_1\Delta_2-\gamma_1\gamma_2)^2+(\Delta_1\gamma_2-\Delta_2\gamma_1)^2\right]^2}  \nonumber\\
&\hspace{4mm}\times \left[(\kappa^2-\gamma_1\gamma_2)^2-\Delta_1^2\Delta_2^2-(\Delta_1^2\gamma_2^2-\Delta_2^2\gamma_1^2)\right] \bigg]. \nonumber\\
\end{align}
\begin{eqnarray}\label{Parameters of reduced mechanical motion equations 2}
\eta_1&=&\frac{g_{om}\left[\gamma_2^2\epsilon_1^2+(\Delta_2\epsilon_1-\kappa\epsilon_2)^2\right]}
{(\kappa^2-\Delta_1\Delta_2-\gamma_1\gamma_2)^2+(\Delta_1\gamma_2-\Delta_2\gamma_1)^2},  \nonumber\\
\eta_2&=&\frac{g_{om}\left[\gamma_1^2\epsilon_2^2+(\kappa\epsilon_1-\Delta_1\epsilon_2)^2\right]}
{(\kappa^2-\Delta_1\Delta_2-\gamma_1\gamma_2)^2+(\Delta_1\gamma_2-\Delta_2\gamma_1)^2}.
\end{eqnarray}
%\begin{align}
%&\kappa_{\rm{mech}} =4g_{om}^2\kappa\times \nonumber\\ &\hspace{4mm}\bigg[\frac{\kappa\gamma_{1ex}\epsilon_1^2\left[\Delta_2(\kappa^2-\Delta_1\Delta_2-2\gamma_1\gamma_2)+\Delta_1\gamma_2^2\right]}
%{\left[(\kappa^2-\Delta_1\Delta_2-\gamma_1\gamma_2)^2+(\Delta_1\gamma_2-\Delta_2\gamma_1)^2\right]^2}  \nonumber\\
%&\hspace{4mm}+\frac{+\kappa\gamma_{2ex}\epsilon_2^2\left[\Delta_1(\kappa^2-\Delta_1\Delta_2)-\Delta_2\gamma_1^2\right]}
%{\left[(\kappa^2-\Delta_1\Delta_2-\gamma_1\gamma_2)^2+(\Delta_1\gamma_2-\Delta_2\gamma_1)^2\right]^2} \nonumber\\
%&\hspace{4mm}-\frac{\sqrt{\gamma_{1ex}\gamma_{2ex}}
%\epsilon_1\epsilon_2}
%{\left[(\kappa^2-\Delta_1\Delta_2-\gamma_1\gamma_2)^2+(\Delta_1\gamma_2-\Delta_2\gamma_1)^2\right]^2}  \nonumber\\
%&\hspace{4mm}\times \left[(\kappa^2-\gamma_1\gamma_2)^2-\Delta_1^2\Delta_2^2-(\Delta_1^2\gamma_2^2-\Delta_2^2\gamma_1^2)\right] \bigg]. \nonumber\\
%\end{align}

\section{The optomechanics-induced effective mechanical frequency shifts and mechanical coupling}\label{Appendix: effective frequency shifts and coupling}
\begin{figure}[h]
\centerline{
\includegraphics[width=8.6 cm,clip]{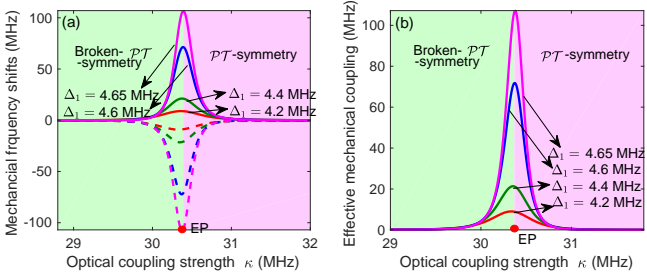}}
\caption{(Color online) (a) The optomechanics-induced mechanical
frequency shifts $\delta\Omega_{1,2}$ versus the optical coupling
strength $\kappa$ in the broken-$\mathcal{PT}$-symmetric regime
(light green area) and $\mathcal{PT}$-symmetric regime (pink
area). Here, we fix $\Delta_2=5$ MHz and plot the curves of
$\delta\Omega_{1,2}$ for different $\Delta_1$. The solid (dashed)
curves denote the curves of the mechanical frequency shift
$\delta\Omega_1$ ($\delta\Omega_2$) with different $\Delta_1$.
(b) The effective mechanical coupling strength $\kappa_{\rm mech}$
between the two mechanical modes versus the optical coupling strength $\kappa$.
}\label{Fig9}
\end{figure}

Let us assume that the gain and loss are well-balanced such that
$\gamma_1=\gamma_2\equiv \gamma$ and consider the critical
coupling case such that $\gamma_{1ex}=\gamma_{2ex}=\gamma/2$. When
$g_{om}\ll|\Delta_1-\Delta_2|\ll \kappa, \gamma$ (or Eq.~\ref{Weaker condition 2})
and $\epsilon_1=\epsilon_2 \equiv \epsilon$, the two mechanical frequency shifts
$\delta\Omega_{1,2}$ in Eq.~\ref{Parameters of reduced mechanical motion equations} can be simplified as
\begin{equation}\label{Mechanical frquency shift in the vicinity of exceptional point}
\delta\Omega_{1}=-\delta\Omega_2\approx\frac{g_{om}^2\Delta_{-}(\gamma^2+\kappa^2)^2\gamma\epsilon^2}
{\left[(\kappa^2-\gamma^2)^2+\gamma^2\Delta_{-}^2 \right]^2},
\end{equation}
we show the optomechanics-induced mechanical frequency shifts
$\delta\Omega_{1,2}$ in Fig.~\ref{Fig9} (a). When the system is
far away from EP, the mechanical frequency shifts $\delta\Omega_1$
(red solid line) and $\delta\Omega_2$ (red dashed line) are very
small, and can be omitted in comparison to the mechanical frequencies
$\Omega_{1,2}$. However, both frequency shifts $\delta\Omega_1$
and $\delta\Omega_2$ will be greatly amplified in the vicinity of
EP, which will modify the mechanical frequencies $\Omega_{1,2}$
such that $\Omega_1+\delta\Omega_1=\Omega_2+\delta\Omega_2$. As
shown in Fig.~\ref{Fig9} (a), in the $\mathcal{PT}$-symmetric
regime of the optical modes, these mechanical frequency shifts are enhanced with the
decrease of the optical coupling strength $\kappa$, which means
that smaller coupling strength $\kappa$ between two optical modes is better for
synchronization. In addition, as shown in Fig.~\ref{Fig9} (a), the
difference between the detuning frequencies of the two optical
modes, i.e., $|\Delta_2-\Delta_1|$, significantly influences the
amplification effects of the mechanical frequency shifts
$\delta\Omega_{1,2}$ when the system is around EP. By fixing
$\Delta_2=5$ MHz, we plot the curves of $\delta\Omega_{1,2}$ for
different $\Delta_1$. We can see that the mechanical frequency
shifts $\delta\Omega_{1,2}$ are greatly enhanced with the decrease
of $|\Delta_2-\Delta_1|$ in the vicinity of EP.

Under the same assumptions, the strength of the effective mechanical
coupling in Eq.~\ref{Parameters of reduced mechanical motion equations}
can be simplified as
\begin{equation}\label{Effective mechanical coupling in the vicinity of exceptional point}
\kappa_{\rm mech} \approx
\frac{4g_{om}^2\Delta_{-}\kappa^2\gamma^3\epsilon^2}
{\left[ (\kappa^2-\gamma^2)^2+\gamma^2\Delta_{-}^2 \right]^2},
\end{equation}
and thus the effective mechanical coupling will be greatly
amplified in the vicinity of EP.
We then plot the curves of the effective mechanical coupling
strength $\kappa_{\rm mech}$ versus the optical coupling strength
$\kappa$ in Fig.~\ref{Fig9} (b). Here we also fix $\Delta_2=5$ MHz
and tune the detuning frequency $\Delta_1$. It can be seen that
the effective mechanical coupling strength $\kappa_{\rm{mech}}$ is
significantly enhanced in the vicinity of EP. Therefore, in the
$\mathcal{PT}$-symmetric regime of the optical modes, weaker optical coupling strength
leads to stronger effective mechanical coupling strength, and thus
may be helpful for the synchronization between the two mechanical
modes. It is also shown that the degree of amplification of
$\kappa_{\rm{mech}}$ is extensively enhanced with the decreasing of
$|\Delta_2-\Delta_1|$ in the vicinity of EP.

\section{The influence of the effective mechanical coupling on synchronization}\label{The influence of the effective mechanical coupling on synchronization}
In this part we discuss the positive effect of the enhancement
of the effective mechanical coupling $\kappa_{\rm{mech}}$ on the synchronization
between mechanical modes, i.e., the stronger the $\kappa_{\rm{mech}}$ is,
the easier the synchronization is. For simplicity and clarity, we re-express
the dynamical equation in Eq.~(\ref{Reduced mechanical motion equations}) by using
the differential operator format as follows
\begin{eqnarray}
\left[\mathcal{D}+(\Gamma_{m1}+i(\Omega_1+\delta\Omega_1)) \right]\beta_1
+\kappa_{\rm{mech}}\beta_2&=&-i\eta_1, \nonumber\\
\kappa_{\rm{mech}}\beta_1 +
\left[ \mathcal{D}+(\Gamma_{m2}+i(\Omega_2+\delta\Omega_2)) \right]\beta_2&=&-i\eta_2, \nonumber
\end{eqnarray}
where $\mathcal{D}$ represents the differential operator. By eliminating
the degree of freedom of $\beta_2$, we can derive the dynamical equation
of $\beta_1$, and then obtain the characteristic equation of this coupled
system as follows
\begin{align}
&\lambda^2+\left[\Gamma_{m1}+\Gamma_{m2}+i(\Omega_1+\delta\Omega_1+\Omega_2+\delta\Omega_2) \right]\lambda  \nonumber\\
&+\left[ \Gamma_{m1}+i(\Omega_1+\delta\Omega_1) \right]
\left[ \Gamma_{m2}+i(\Omega_2+\delta\Omega_2) \right]
-\kappa_{\rm{mech}}^2=0. \nonumber
\end{align}
By considering $\Gamma_{m1}=\Gamma_{m2}=\Gamma_m$, the roots of this characteristic
equation can be expressed as
\begin{eqnarray}\label{Characteristic equation of the mechanical system}
\lambda_{+}&=&-\Gamma_m-i\Omega_{Ave+}
+i\sqrt{\Omega_{Ave-}^2-\kappa_{\rm{mech}}^2}, \nonumber\\
\lambda_{-}&=&-\Gamma_m-i\Omega_{Ave+}
-i\sqrt{\Omega_{Ave-}^2-\kappa_{\rm{mech}}^2},
\end{eqnarray}
where
\begin{eqnarray}
\Omega_{Ave+}&=&\frac{\Omega_1+\delta\Omega_1+\Omega_2+\delta\Omega_2}{2}, \nonumber\\
\Omega_{Ave-}&=&\frac{\Omega_1+\delta\Omega_1-\Omega_2-\delta\Omega_2}{2}. \nonumber
\end{eqnarray}
It can be easily seen that in the weak coupling regime such that $\kappa_{\rm{mech}}<\Omega_{Ave-}$,
the vibration frequencies of the mechanical modes $\beta_{1,2}$ are close to each other
with the increase of the effective coupling strength $\kappa_{\rm{mech}}$, which means that the degree
of synchronization between the two mechanical modes increases with the increase of $\kappa_{\rm{mech}}$.
At the critical point such that $\kappa_{\rm{mech}}=\Omega_{Ave-}$, the two oscillators
will have the same vibration frequency $\Omega_{Ave+}$, which means that these two mechanical
modes are with frequency synchronization, i.e., the frequencies of the two mechanical modes
are equal to each other.
It is shown that a stronger effective mechanical coupling strength can improve the
degree of the synchronization between mechanical modes in our system, and finally leads to
the frequency synchronization when the effective mechanical coupling is strong enough.

In addition, in the weak coupling regime, the Eq.~(\ref{Characteristic equation of the mechanical system})
can also be re-expressed as
\begin{eqnarray}\label{Characteristic equation of the mechanical system 2}
\lambda_{+}&=&\Gamma_m - i\left( \Omega_2+\delta\Omega_2-\delta\Omega_{\rm{coup}} \right)   \nonumber\\
\lambda_{-}&=&\Gamma_m - i\left( \Omega_1+\delta\Omega_1+\delta\Omega_{\rm{coup}} \right),
\end{eqnarray}
where
\begin{eqnarray}
\delta\Omega_{\rm{coup}} &=& \frac{\Omega_2+\delta\Omega_2-\Omega_1-\delta\Omega_1}{2} \nonumber\\
&&-\sqrt{\left( \frac{\Omega_2+\delta\Omega_2-\Omega_1-\delta\Omega_1}{2} \right)^2 - \kappa_{\rm{mech}}^2} \nonumber\\
\end{eqnarray}
is induced by the effective mechanical coupling strength $\kappa_{\rm{mech}}$. It can be seen
in Eq.~(\ref{Characteristic equation of the mechanical system 2}) that both
optomechanics-induced mechanical frequency shift $\delta\Omega_i$ and
effective mechanical coupling $\kappa_{\rm{mech}}$ can lead to frequency shifts of the
two mechanical modes, and thus contribute to the synchronization together.

\section{The enhancement of the effective optomechanical interaction}\label{The enhancement of the effective optomechanical interaction}
In our $\mathcal{PT}$-symmetric optomechanical system, there exists an enhancement
of the effective optomechanical interaction due to the topological-singularity-induced
amplification of optomechanical nonlinearity in the vicinity of the exceptional point \cite{Jinghui,JZhang}.
This enhanced optomechanical interaction then leads to the amplifications
of the optomechanics-induced mechanical frequency shifts $\delta\Omega_{1,2}$ and
the effective mechanical coupling strength $\kappa_{\rm{mech}}$.
Since both the optomechanics-induced mechanical frequency shifts and the effective
mechanical coupling can change the frequency of the two mechanical modes, thus the
synchronization between far-off-resonant mechanical modes can be realized with
sufficiently large optomechanical interaction strength. In the $\mathcal{PT}$-symmetric regime,
the system approaches to the exceptional point with the decrease of optical coupling strength $\kappa$,
which results in an enhancement of the optomechanical coupling and thus compensates
the reduction of the optical coupling strength. In the following part of
this subsection, we will discuss this enhanced effective optomechanical
interaction in our $\mathcal{PT}$-symmetric optomechanical system.

\begin{figure}[h]
\centerline{
\includegraphics[width=6 cm,clip]{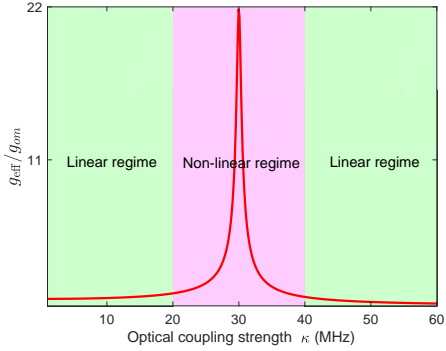}}
\caption{(Color online) Effective optomechanical coupling strength $g_{\rm{eff}}$ versus the optical coupling
strength $\kappa$. In the green area, the system is far away from EP, and the effective
optomechanical coupling strength $g_{\rm{eff}}$ is linearly dependent on $\kappa$.
In the pink area, the system is in the vicinity of EP, and in this case, $g_{\rm{eff}}$ changes
nonlinearly with $\kappa$.}\label{Fig10}
\end{figure}

In our optomechanical system, the interaction Hamiltonian between
optical modes and mechanical modes can be expressed as
\begin{eqnarray}\label{Interaction Hamiltonian}
H_{int}=g_{om}a_1^{\dagger} a_1(b_1^{\dagger}+b_1)+g_{om}a_2^{\dagger}
a_2(b_2^{\dagger}+b_2),
\end{eqnarray}
where $a_1$ ($a_2$) and $b_1$ ($b_2$) represent the annihilation operator of
the optical mode and mechanical mode in the active (passive) resonator, respectively,
and $g_{om}$ is the optomechanical coupling strength. If we re-write this
interaction Hamiltonian $H_{int}$ in the optical supermodes picture, then the effective optomechanical
coupling strength $g_{\rm{eff}}$ between optical supermodes and mechanical
modes can be expressed as
\begin{eqnarray}\label{Effective optomechanical coupling}
g_{\rm{eff}}\approx \frac{g_{om}}{2}\frac{\gamma^2+\sqrt{(\kappa^2-\gamma^2)^2+\gamma^2\Delta_{-}^2}}
{\sqrt{(\kappa^2-\gamma^2)^2+\gamma^2\Delta_{-}^2}}.
\end{eqnarray}
Since $\Delta_{-}=|\Delta_2-\Delta_1|\ll \kappa, \gamma$, the effective optomechanical
coupling strength $g_{\rm{eff}}$ can be greatly amplified in the vicinity of EP when $\kappa \to \gamma$.
This means that in this case the effective optomechanical coupling strength $g_{\rm{eff}}$
can be greatly enhanced. Given the parameters, we can obtain the simulation results
of the effective optomechanical coupling strength $g_{\rm{eff}}$ versus the optical coupling strength $\kappa$, as shown in Fig.~\ref{Fig10}.
When the optical coupling strength $\kappa$ is far away from the exceptional point, i.e., in the green area in Fig.~\ref{Fig10},
the effective optomechanical coupling strength changes linearly with the optical coupling strength $\kappa$. However,
in the pink area, $g_{\rm{eff}}$ increases very fast when the system approaches to
EP, which means that in this regime the optomechanical interaction can be greatly
amplified. In addition, by comparing Eq.~(\ref{Effective optomechanical coupling}) with
Eq.~(\ref{Mechanical frquency shift in the vicinity of exceptional point})
and Eq.~(\ref{Effective mechanical coupling in the vicinity of exceptional point}),
we can find that $|\delta\Omega_{1,2}|\propto g_{\rm{eff}}^4|f_1(\kappa,\gamma,\epsilon,g_{om},\Delta_{-})|$, and
$\kappa_{\rm{mech}}\propto g_{\rm{eff}}^4 |f_2(\kappa,\gamma,\epsilon,g_{om},\Delta_{-})|$, which means that the enhanced
optomechanical coupling strength can lead to improvements of the optomechanics-induced mechanical
frequency shifts and the effective mechanical coupling in the vicinity of EP.

\section{The difference between active $\mathcal{PT}$-symmetric system and passive system with EP for synchronization}
Based on the previous discussion, we know that in the discussed gain-loss balanced
$\mathcal{PT}$-symmetric optomechanical system, there exists amplifications
of the optomechanics-induced mechanical frequency shifts and effective mechanical coupling strength
in the vicinity of exceptional point. However,
if this $\mathcal{PT}$-symmetric system is replaced by a passive coupled system with an exceptional point,
i.e., the active resonator in the discussed $\mathcal{PT}$-symmetric system is replaced by
a passive resonator, the two far-detuned mechanical modes in this system cannot
synchronize with each other. To show this, we can easily obtain the dynamical equations
of the system by replacing the optical damping $\gamma_1$ in
Eq.~(\ref{Dynamical Equations of the optomechanical system}) with $-\gamma_1$
\begin{eqnarray}\label{Dynamical Equations of the non-PT-symmetric system}
\dot{\alpha}_1&=&(-\gamma_1-i\Delta_1)\alpha_1-i \kappa \alpha_2-i g_{om}\alpha_1(\beta_1+\beta^*_1) \nonumber\\
&&+\sqrt{2\gamma_{1ex}}\epsilon_1,\nonumber\\
\dot{\alpha}_2&=&(-\gamma_2-i\Delta_2)\alpha_2-i \kappa \alpha_1-i g_{om}\alpha_2(\beta_2+\beta^*_2)  \nonumber\\
&&+\sqrt{2\gamma_{2ex}}\epsilon_2,\nonumber\\
\dot{\beta}_1&=&-(\Gamma_{m1}+i\Omega_1)\beta_1-i g_{om} |\alpha_1|^2,\nonumber\\
\dot{\beta}_2&=&-(\Gamma_{m2}+i\Omega_2)\beta_2-i g_{om}
|\alpha_2|^2.
\end{eqnarray}
Under the assumptions that $\sqrt{2\gamma_{1ex}}\epsilon_1=\sqrt{2\gamma_{2ex}}\epsilon_2=\epsilon$,
and $\Delta_{1,2}, |\Delta_2-\Delta_1|\ll \kappa, \gamma_{1,2}$,  the optomechanics-induced
mechanical frequency shifts $\delta\Omega_{1,2}$ and the effective mechanical coupling
$\kappa_{\rm{mech}}$ can be approximately expressed as
\begin{eqnarray}
\delta\Omega_1&=&-\delta\Omega_2\approx g_{om}^{2}\frac{\Delta_{-}\epsilon^2}
{\left[(\kappa^2+\gamma_1\gamma_2)+\Delta_{+}^{2}\right]^2}, \nonumber\\
\kappa_{\rm{mech}}&\approx&2g_{om}^{2}\frac{\kappa\epsilon^2}
{\left[(\kappa^2+\gamma_1\gamma_2)+\Delta_{+}^{2}\right]^2},
\end{eqnarray}
where $\Delta_{+}=(\Delta_1+\Delta_2)/2$ and $\Delta_{-}=\Delta_2-\Delta_1$.
As $\Delta_{-}\ll \kappa, \gamma_{1,2}$ and $g_{om}$ is very tiny,
$\delta\Omega_{1,2}$ and $\kappa_{\rm{mech}}$ are very small. This implies that
in this passive system with an exceptional point, the amplifications of mechanical
frequency shifts and effective mechanical coupling are not strong enough. Thus
these two mechanical modes with far-off-resonant mechanical frequencies cannot
be synchronized.

\begin{figure}[h]
\centerline{
\includegraphics[width=8.8 cm,clip]{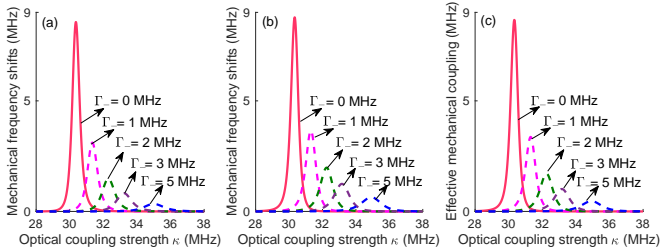}}
\caption{(Color online) (a) Optomechanics-induced mechanical frequency shifts
$\delta\Omega_1$ versus the optical coupling strength $\kappa$ with different
$\Gamma_{-}$. The solid curve denotes the case that gain and loss are balanced, i.e.,
$\Gamma_{-}=0$. It is shown that the amplification effects of $\delta\Omega_1$
are suppressed with the increase of $\Gamma_{-}$. (b) Corresponding to
the optomechanics-induced mechanical frequency shifts $-\delta\Omega_2$
versus the optical coupling strength $\kappa$ with different
$\Gamma_{-}$. (c) Effective mechanical coupling
$\kappa_{\rm{mech}}$ between the two mechanical modes versus the optical coupling
strength $\kappa$ with different $\Gamma_{-}$. It is shown that
the amplification effects of $\kappa_{\rm{mech}}$ are also suppressed
with the increase of $\Gamma_{-}$}\label{Fig11}
\end{figure}

In addition, if the balance between gain and loss is broken in our $\mathcal{PT}$-symmetric
system, i.e., $\Gamma_{-}=|\gamma_1-\gamma_2|/2\ne 0$, the synchronization
between the two mechanical modes will be suppressed. In fact,
when the balance between gain and loss is broken, the mechanical frequency shifts
$\delta\Omega_{1,2}$ and the effective mechanical coupling $\kappa_{\rm{mech}}$ can
be expressed as
\begin{align}
&\delta\Omega_1 \approx 2g_{om}^{2}\frac{\Delta_{-}(\kappa^2+\gamma_2^2)^2\epsilon^2}
{\left[(\kappa^2-\gamma_1\gamma_2)^2+(\gamma_1+\gamma_2)^2\Delta_{-}^2/4+\Gamma_{-}^2\Delta_{+}^2 \right]^2}, \nonumber \\
&\delta\Omega_2 \approx 2g_{om}^{2}\frac{\Delta_{-}(\kappa^2+\gamma_1^2)^2\epsilon^2}
{\left[(\kappa^2-\gamma_1\gamma_2)^2+(\gamma_1+\gamma_2)^2\Delta_{-}^2/4+\Gamma_{-}^2\Delta_{+}^2 \right]^2}, \nonumber \\
&\kappa_{\rm{mech}} \approx 4g_{om}^2\frac{\Delta_{-}\kappa^2\gamma_1\gamma_2\epsilon^2}
{\left[(\kappa^2-\gamma_1\gamma_2)^2+(\gamma_1+\gamma_2)^2\Delta_{-}^2/4+\Gamma_{-}^2\Delta_{+}^2 \right]^2}. \nonumber \\
\end{align}
Therefore, with the increase of $\Gamma_{-}$, the amplification effects
of the mechanical frequency shifts and the effective mechanical coupling strength will be suppressed.
We show the mechanical frequency shifts $\delta\Omega_{1,2}$
and the effective mechanical coupling strength $\kappa_{\rm{mech}}$ with different $\Gamma_{-}$
in Figs.~\ref{Fig11} (a), (b), and (c), respectively.
It can be clearly seen that the amplifications of the mechanical frequency shifts and the effective
mechanical coupling strength are seriously suppressed when $\Gamma_{-}$ is large,
thus the synchronization between the two mechanical modes with far-off-resonant
cannot be realized.

\section{Derivation of the normalized correlation function $R$}\label{Derivation of the normalized correlation function}

To simplify our discussions, we redefine four variables
\begin{eqnarray}
\xi_1&=&x_1, \quad \xi_2=\dot{\xi}_1, \nonumber \\
\xi_3&=&x_2, \quad \xi_4=\dot{\xi}_3,
\end{eqnarray}
thus the Langevin equation of the mechanical modes (Eq.~\ref{The reduced Langevin equation with Brownian noises})
can be re-expressed as
\begin{eqnarray}\label{Convenient edition of Langevin equation}
\left[\begin{matrix}
\dot{\xi}_1(t) \\ \dot{\xi}_2(t) \\ \dot{\xi}_3(t) \\ \dot{\xi}_4(t)
\end{matrix}\right]&=&-
\left[\begin{matrix}
0 & -1 & 0 & 0\\  \tilde{\Omega}_1^2 & 2\Gamma_m & \kappa_{\rm{mech}} & 0\\
0 & 0 & 0 & -1\\ \kappa_{\rm{mech}} & 0 & \tilde{\Omega}_2^2 & 2\Gamma_m
\end{matrix}\right]
\left[\begin{matrix}
\xi_1(t) \\ \xi_2(t) \\ \xi_3(t) \\ \xi_4(t)
\end{matrix}\right]+
\left[\begin{matrix}
\Gamma_1 \\ \Gamma_2 \\ \Gamma_3 \\ \Gamma_4
\end{matrix}\right] \nonumber \\
&=&
-A
\left[\begin{matrix}
\xi_1(t) \\ \xi_2(t) \\ \xi_3(t) \\ \xi_4(t)
\end{matrix}\right]+
\left[\begin{matrix}
\Gamma_1 \\ \Gamma_2 \\ \Gamma_3 \\ \Gamma_4
\end{matrix}\right],
\end{eqnarray}
where $\Gamma_1=\Gamma_3=0$; $\Gamma_2=\Gamma_{\rm{noise1}}$;
and $\Gamma_4=\Gamma_{\rm{noise2}}$.
The solution of the above equation can be expressed as
\begin{eqnarray}\label{General solution}
\xi_i(t)&=&
\sum\limits_{k=1}^4 G_{ik}(t) z_k +
\sum\limits_{k=1}^4
\int_0^t G_{ik}(t')\Gamma_k(t-t')dt',\nonumber\\
\end{eqnarray}
where matrix $G=(G_{ij})=\exp(-At)$, and $z_i$ represent the initial values of
the variables $\xi_i$.
As we consider small time $t$, the matrix $G$ can be
approximately expressed as
\begin{eqnarray}
G(t)&=&e^{-At} \nonumber\\
&\approx&I-At \nonumber\\
&\approx&
\left[\begin{matrix}
1 & t & 0 & 0 \\
-\tilde \Omega_1^2 t & 1-2\Gamma_m t & -\kappa_{\rm{mech}}t & 0 \\
0 & 0 & 1 & t \\
-\kappa_{\rm{mech}}t & 0 & -\tilde \Omega_2^2 t & 1-2\Gamma_m t
\end{matrix}\right], \nonumber\\
\end{eqnarray}
thus the solution of $\xi_1 (t)$ in Eq.~(\ref{General solution}) can be approximately
expressed as
\begin{eqnarray}\label{Solutions_of_xi1}
\xi_1(t)&=&G_{11}z_1+G_{12}z_2+G_{13}z_3+G_{14}z_4 \nonumber\\
        &&+\int_{0}^{t}{G_{11}(t')\Gamma_1(t-t') + G_{12}(t')\Gamma_2(t-t')} \nonumber\\
        &&+G_{13}(t')\Gamma_3(t-t')+G_{14}(t')\Gamma_4(t-t')dt'  \nonumber\\
        &=&z_1+t z_2 +\int_{0}^{t} {t'\Gamma_2(t-t')dt'}.
\end{eqnarray}
Similarly, other solutions in Eq.~(\ref{General solution}) can be approximately
expressed as
\begin{eqnarray}\label{Solutions_of_xi234}
\xi_2(t)&=&-\left( \kappa_{\rm{mech}}z_3+ \tilde \Omega_1^2 z_1 \right)t+
\left( 1-2\Gamma_m t \right)z_2 \nonumber\\
&&+ \int_{0}^{t}{\left(1-2\Gamma_m t' \right)\Gamma_2(t-t')dt'}, \nonumber\\
\xi_3(t)&=&z_3+z_4 t +\int_{0}^{t}{t' \Gamma_4(t-t')dt'}, \nonumber\\
\xi_4(t)&=&-\left( \kappa_{\rm{mech}}z_1+\tilde \Omega_2^2 z_3 \right)t+
\left( 1-2\Gamma_m t \right)z_4 \nonumber\\
&&+ \int_{0}^{t}{\left( 1-2\Gamma_m t' \right)\Gamma_4(t-t')dt'}.
\end{eqnarray}
We then calculate the correlation functions as
\begin{eqnarray}\label{Two time correlation}
R_{ij}(\tau,t)=\left\langle \xi_i(t+\tau)\xi_j(t) \right\rangle,
\end{eqnarray}
where $\left\langle\; \cdot \; \right\rangle$ is
the ensemble average over the stochastic noises.
Based on the regression theorem~\cite{Risken},
we know that the correlation functions $R_{ij}(\tau,t)$
can be reduced to
\begin{eqnarray}\label{Reduced correlation function}
R_{ij}(\tau,t) &=&\sum\limits_{k=1}^4 G_{ik}(\tau)
\left\langle \xi_k(t) \xi_j(t) \right\rangle, \quad 0\le \tau.
\end{eqnarray}
By substituting the solutions of $\xi_{i}(t)$ as shown in Eqs.~(\ref{Solutions_of_xi1})
and (\ref{Solutions_of_xi234}) into the correlation functions $R_{ij}(\tau,t)$ (Eq.~(\ref{Reduced correlation function})),
the three correlation functions $R_{13}(\tau,t)$, $R_{11}(0,t)$, and $R_{33}(0,t)$
can be expressed as
\begin{align}
&R_{13}(\tau,t)=(z_1+t z_2)(z_3+z_4 t) +\frac{q}{3}t^3 \nonumber\\
&\hspace{6mm}+\tau \Bigg[ -\left( \kappa_{\rm{mech}}z_3 + \tilde \Omega_1^2 z_1 \right)(z_3+z_4 t)t\nonumber\\
&\hspace{6mm}+(1-2\Gamma_m t)(z_3+z_4 t)z_2 +q\left( \frac{1}{2}t^2-\frac{2}{3}\Gamma_m t^3 \right) \Bigg], \nonumber\\
&R_{11}(0,t)=(z_1+z_2 t)^2+\frac{q}{3}t^3,\nonumber\\
&R_{33}(0,t)=(z_3+z_4 t)^2+\frac{q}{3}t^3.
\end{align}
For simplicity we assume that the system is stationary at the initial time,
i.e., $z_2=z_4=0$, and consider the case that $z_1=1/\tilde\Omega_1^2, z_3=1/\kappa_{\rm{mech}}$,
thus the normalized correlation function between the two mechanical modes
can be expressed as
\begin{eqnarray}\label{Normalized correlation function2}
R(\tau,t)&=&\frac{|R_{13}(\tau,t)|}{\sqrt{R_{11}(0,t)}\sqrt{R_{33}(0,t)}} \nonumber\\
&=&\frac{\left| 1-2\tilde\Omega_1^2\tau t+q\frac{\kappa_{\rm{mech}}\tilde\Omega_1^2}{2}\tau t^2
+q\frac{\kappa_{\rm{mech}}\tilde\Omega_1^2}{3} \tau t^3 \right|}{\sqrt{1+\frac{q}{3}\kappa_{\rm{mech}}^2  t^3}
\sqrt{1+\frac{q}{3}\tilde\Omega_1^4 t^3}} \nonumber \\
&\approx& 1-2\tilde\Omega_1^2\tau t + \frac{q}{2} \kappa_{\rm{mech}}\tilde\Omega_1^2\tau t^2    +\frac{q}{3} \kappa_{\rm{mech}}\tilde\Omega_1^2 \tau t^3\nonumber\\
&\approx& 1-2\tilde\Omega_1^2\tau t+2\Gamma_m k T\kappa_{\rm{mech}}\tilde\Omega_1^2\tau t^2\nonumber\\
&&+\frac{4}{3}\Gamma_m k T\kappa_{\rm{mech}}\tilde\Omega_1^2 \tau t^3.
\end{eqnarray}

%\section*{APPENDIX: CHAOTIC SYNCHRONIZATION OF COLPITTS OSCILLATOR CIRCUITS}\label{Chaotic synchronization of colpitts oscillator circuits}

\end{document}